\documentclass[12pt]{elsart}
\usepackage{graphicx,amssymb,amsmath,psfrag}
\usepackage[a4paper]{geometry}
\usepackage{ulem}
\usepackage{color}
\usepackage{soul}
\sethlcolor{yellow}

\journal{Nuclear Physics B}

\begin{document}
\begin{frontmatter}
  \title{Percolation of Vortices in the 3D Abelian Lattice Higgs Model}
  \author[Leipz]{Sandro Wenzel}, \author[Leipz]{Elmar Bittner},
  \author[Leipz]{Wolfhard Janke}, and \author[Berlin]{Adriaan
  M.J. Schakel} \address[Leipz]{Institut f\"ur Theoretische Physik and
  Centre for Theoretical Sciences (NTZ), Universit\"at Leipzig, Postfach
  100 920, D-04009 Leipzig, Germany} \address[Berlin]{Institut f\"ur
  Theoretische Physik, Freie Universit\"at Berlin, Arnimallee 14,
  D-14195 Berlin, Germany}

\begin{abstract}
The compact Abelian Higgs model is simulated on a cubic lattice where it
possesses vortex lines and pointlike magnetic monopoles as topological
defects. The focus of this high-precision Monte Carlo study is on the
vortex network, which is investigated by means of percolation
observables.  In the region of the phase diagram where the Higgs and
confinement phases are separated by a first-order transition, it is
shown that the vortices percolate right at the phase boundary, and that
the first-order nature of the transition is reflected by the network.
In the crossover region, where the phase boundary ceases to be
first order, the vortices are shown to still percolate.  In contrast to
other observables, the percolation observables show finite-size scaling.
The exponents characterizing the critical behavior of the vortices in
this region are shown to fall in the random percolation universality
class.
\end{abstract}


\end{frontmatter}

\section{Introduction}
\begin{figure}
\centering 
\psfrag{c}{confined}
\psfrag{h}{Higgs}
\psfrag{n}{$0$}
\psfrag{k}{$\kappa$}
\psfrag{l}{$\lambda$}
\psfrag{lc}{$\lambda_\mathrm{c}$}
\includegraphics[width=0.65\textwidth]{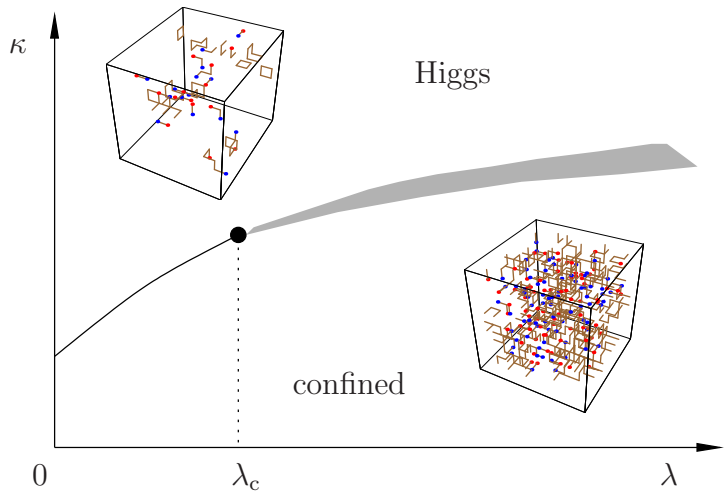}
\caption{\label{fig:phase_diagram} Sketch of the phase diagram of the
  Abelian Lattice Higgs Model as a function of the Higgs self-coupling
  parameter $\lambda$ and the hopping parameter $\kappa$ at fixed
  inverse gauge coupling $\beta$.  For $\lambda<\lambda_c$ the
  transition from the confined to the Higgs phase is of first-order. For
  $\lambda>\lambda_c$ ordinary observables show no singular behavior and
  signal a crossover between the two ground states indicated by the dark
  region emanating from the point where the first-order transition line
  terminates.}
\end{figure}

The Abelian Higgs model with a compact gauge field formulated on a
three-dimensional (3D) lattice possesses an intriguing phase structure
\cite{FradkinShenker,Banks:1979fi,Bowler:1981cj}.  In addition to the
Higgs state where the photon acquires a mass, it exhibits a state in
which electric charges are confined.  The richness of the model, which
serves as a toy model for quark confinement, stems from the presence of
two types of topological excitations, viz.\ vortex lines and magnetic
monopoles.  The latter are point defects in three dimensions which arise
because of the compactness of the U(1) gauge group.  In the pure 3D
compact Abelian gauge theory, monopoles are known to form a plasma that
physically causes confinement of electric charges for all values of the
inverse gauge coupling $\beta$ \cite{Polyakov}.  Being the only
parameter present, the pure model therefore possesses only a confinement
phase.  The coupling to the scalar theory preserves this confinement
state and gives in addition rise to a Higgs state.  For sufficiently
small values of the Higgs self-coupling parameter $\lambda$, the two
ground states are separated by a first-order transition
\cite{Munehisa:1985rb,Obodi:1985uu} as sketched in
Fig.~\ref{fig:phase_diagram}. For $\lambda$ larger than a critical value
$\lambda_\mathrm{c}$, which depends on the value of the gauge coupling,
the two states are no longer separated by a phase boundary across which
thermodynamic observables become singular, as was first shown by Fradkin
and Shenker \cite{FradkinShenker} in the limit $\lambda \to \infty$
where fluctuations in the amplitude of the Higgs field become completely
suppressed.  In other words, it is always possible to cross over from
one ground state to the other without encountering a thermodynamic
singularity.  Because of this, the Higgs and confinement states were
thought to constitute a single phase, despite profound differences in
physical properties.

This conclusion is supported by symmetry arguments
\cite{KovnerRosenstein}.  The relevant global symmetry group of the
compact 3D Abelian Higgs model (cAHM) is the cyclic group Z$_q$ of $q$
elements, where the integer $q$ denotes the electric charge of the Higgs
field.  For the doubly charged case ($q=2$), the relevant symmetry group
is Z$_2$, which is in agreement with the known result that the model
undergoes a continuous phase transition belonging to the 3D Ising
universality class \cite{FradkinShenker,Bhanot:1981ug}.  For $q=1$, this
argument excludes a phase transition characterized by a local order
parameter in the spirit of Landau because the group $Z_1$, which
consists of only the unit element, cannot be spontaneously broken.
Stated differently, there exist no local order parameter which
distinguishes the Higgs from the confinement state.

In Ref.~\cite{Wenzel:2005nd}, we argued that the phase diagram is more
refined than implied by this picture.  We conjectured that although
analytically connected, the two ground states can be considered as two distinct phases.  
The nature of the phase boundary is intimately connected to the distinct
physical properties of the Higgs and confinement phase. For the latter
phase to confine electric charges, the monopoles must form a plasma.
This in turn can only happen when the line tension of the vortex lines,
or flux tubes, connecting monopoles and antimonopoles vanishes, so that
they are no longer tightly bound in pairs as in the Higgs phase (for
typical configuration plots see the snapshots in
Fig.~\ref{fig:phase_diagram}). Since the vortex line tension is finite
in the Higgs phase and zero in the confinement phase, we argued that the
phase boundary is uniquely defined by the vanishing of the vortex line
tension, irrespective of the order of the phase transition.  The
confinement mechanism operating in the 3D cAHM is essentially the dual
superconductor scenario~\cite{Mandelstam:1974pi,'tHooft:1977hy}.

In addition to open vortex lines, each having a monopole and an
antimonopole at its endpoints, the system also possesses closed vortex
lines.  These are expected to be characterized by the same line tension
as the open lines.  Because of the finite line tension, large vortex
loops are exponentially suppressed in the Higgs phase.  Upon approaching
the phase boundary, the line tension becomes smaller so that the vortex
network can grow larger and the overall line density increases.
Finally, at the phase boundary where the line tension vanishes, vortices
can grow arbitrarily large at no energy cost.  The phase boundary between
the Higgs and confinement phase is therefore expected to be marked by a
proliferation of (open and closed) vortex lines, as was first proposed
by Einhorn and Savit~\cite{einhorn2}.  The vortices proliferate both in
the region where the transition is first-order and in the region where
it is not.  A line along which geometrical objects proliferate, yet
thermodynamic quantities and other local gauge-invariant observables
remain nonsingular has become known as a Kert\'esz line \cite{Kertesz}.
Such a line was first discussed in the context of spin clusters in the
2D Ising model in an applied magnetic field.

The same deconfinement transition driven by the proliferation of
vortices with a phase boundary consisting of a first-order line which
ends in a critical point and then continues as a Kert\'esz line was
originally proposed by Langfeld for the SU(2) counterpart of the Abelian
Higgs model defined on a 4D lattice \cite{Langfeld:2002ic}.  The
relevant vortices, forming surfaces in 4D, are center vortices which
carry a flux related to the nontrivial element of the Z$_2$ center of
the SU(2) gauge group.  The presence of the Kert\'esz line in this
non-Abelian model has been further numerically investigated and
confirmed in Ref.~\cite{Bertle:2003pj}.  While the concept of a
Kert\'esz line was originally introduced in the context of lattice gauge
theories in Ref.~\cite{satz-2001-681}, the interpretation of a
deconfinement transition as driven by percolating vortices was
previously put forward in the context of the SU(2) lattice Higgs model
\cite{Chernodub:1998wh,Chernodub:1998yv,Bertle:2002mm}, and the pure
SU(2) lattice gauge theory
\cite{Langfeld:1998cz,Engelhardt:1999fd,Langfeld:2003zi}.

The purpose of this Monte Carlo study is to investigate the vortex
proliferation scenario suggested in
Refs.~\cite{Wenzel:2005nd,wenzel-2005-LAT2005} by studying the behavior
of the vortex network directly.  Because vortices are geometrical
objects, their analysis is amenable to the methods developed in
percolation theory \cite{StaufferAharony}.  We conjectured in
Refs.~\cite{Wenzel:2005nd,wenzel-2005-LAT2005} that along the Kert\'esz
line, percolation observables have the usual percolation exponents. In
addition, we expect that the vortex network displays discontinuous
behavior in the region where the phase boundary consists of a
first-order transition.  For other recent Monte Carlo studies focusing
on various aspects of the model see
Refs.~\cite{schiller1,feldmann1,takashima-2006-74} and references
therein.

A similar high-precision Monte Carlo study of the behavior of a vortex
network was recently carried out of the 3D XY and the $|\phi|^4$
lattice model \cite{kajantieperc2000,bittner3}, respectively. The latter
model, whose critical temperature on a cubic lattice is known to high
precision, corresponds to taking $\beta \to \infty$ in the Higgs model
so that the gauge fields become completely ordered.  An important
observation made in that study was that the overall vortex line
density behaves similar to the energy, and the associated
susceptibility similar to the specific heat.  The vortex percolation
threshold estimated through finite-size scaling analysis of the
overall line density data was found to be perfectly consistent with
the critical temperature.  However, estimates based on any of the
vortex percolation observables used, while being close to that
temperature, never coincided with it within error bars.  Possibly, the
mismatches are related to the way the vortex networks are traced out,
or to the absence of a stochastic element as present in, for example,
the Fortuin-Kasteleyn definition of spin clusters in the Potts model
\cite{FrotuinKasteleyn}.  In any case, since we use the same
percolation observables and the same (imperfect) rules to trace out
vortex networks, we expect to find a similar mismatch, at least in the
region where the transition is no longer discontinuous.  To remind the
reader of these qualifications, we will sometimes refer to the
estimated percolation threshold as ``apparent percolation threshold''.

The rest of the paper is organized as follows.  In the next section, the
observables used to investigate the model are introduced.  In
Sec.~\ref{sec:Simulation}, the simulation methods as well as the
numerical tools used to analyze the data are discussed.  In that
section, also a new tool to effectively collapse data gathered on
lattices of different size is presented.  Sections~\ref{sec:1st} and
\ref{sec:kertesz} contain the Monte Carlo results in the vicinity of the
phase boundary for the two regions where it consists of a first-order
transition and a Kert\'esz line, respectively.  In
Sec.~\ref{sec:threshold}, the dependence of the location of the
percolation threshold is investigated as a function of the parameters of
the model. Finally, Sec.~\ref{sec:conclusions} contains a discussion of
the Monte Carlo results and our conclusions.

\section{Definitions and Observables}
\label{sec:defs}
The Abelian lattice Higgs model with compact gauge field at the absolute
zero of temperature is defined by the action $S = S_\mathrm{g} +
S_\phi$, with the gauge part
\begin{equation}
\label{Sg} 
  S_\mathrm{g} = \beta\sum_{x,\mu<\nu} \left[ 1-\cos \theta_{\mu \nu}
  (x) \right],
\end{equation} 
where $\beta$ is the inverse gauge coupling parameter, $\beta=1/aq^2$
with $a$ the lattice spacing and $q$ the electric charge of the Higgs
field.  We exclusively consider the case $q=1$.  The doubly charged
Higgs field ($q=2$), which has an even richer topological structure than
the $q=1$ case, has recently been investigated in Ref.~\cite{feldmann2},
where it was found that the monopoles form chains.  The sum in Eq.~(\ref{Sg})
extends over all lattice sites $x$ and lattice directions $\mu$, and
$\theta_{\mu \nu}(x)$ denotes the usual plaquette variable $\theta_{\mu
\nu} (x) = \Delta_\mu \theta_{\nu} (x) - \Delta_\nu\theta_{\mu} (x)$
with the lattice derivative\linebreak $\Delta_\nu \theta_{\mu} (x) \equiv
\theta_{\mu} (x+\nu) - \theta_{\mu} (x)$ and the compact link variable
$-\pi \leq \theta_{\mu} (x) < \pi$.  The matter part $S_\phi$ of the
lattice action is given by
\begin{equation} 
\label{hopping}
S_\phi = -\kappa\sum_{x,\mu}\rho(x) \rho(x+\mu) \cos 
\left[\Delta_\mu\varphi(x) - q \theta_{\mu}(x)\right] 
+ \sum_x \left\{\rho^2(x) + \lambda 
  \left[\rho^2(x)-1\right]^2 \right\},
\end{equation} 
where polar coordinates are chosen to represent the complex Higgs field
$\phi(x) = \rho(x) \mathrm{e}^{\mathrm{i} \varphi(x)}$, with the compact
phase $-\pi \leq \varphi(x) < \pi$.  In Eq.~(\ref{hopping}), $\kappa$ is
the hopping parameter, and $\lambda$ the Higgs self-coupling.  We study
the theory on a cubic lattice, which either is taken to represent a 3D
space or spacetime box, depending on whether one of the dimensions of
the lattice is interpreted as (Euclidean) time.

In addition to measuring field observables such as the total
action or energy $S$, the hopping energy
\begin{equation} 
\label{hop}
E_\mathrm{h} \equiv \frac{1}{L^3} \sum_{x,\mu}\rho(x)
\rho(x+\mu) \cos \left[\Delta_\mu\varphi(x) - q
\theta_{\mu}(x)\right] ,
\end{equation} 
the square Higgs amplitude,
and  coslink energy 
\begin{equation} 
C \equiv - \frac{1}{3L^3} \sum_{x,\mu} \cos
\left[\Delta_\mu\varphi(x) - q \theta_{\mu}(x) \right] ,
\end{equation} 
we especially probe for topological excitations. A gauge invariant vortex
line segment $j_\lambda(x)$ pointing in the $\lambda=1,2,3$ direction is given by
\begin{equation}
j_\lambda(x) = \epsilon_{\lambda\mu\nu} \nabla_\mu l_\nu(x) - n_\lambda
(x) ,
\end{equation}
where $l_\mu(x)$ is the integer-valued field related to the electric
current along the links of the lattice via
\begin{equation} 
        l_\mu(x) = \frac{1}{2\pi} \left\{\nabla_\mu\varphi(x) - q
        \theta_\mu(x) - [\nabla_\mu\varphi(x) - q
        \theta_{\mu}(x)]_{2\pi}\right\},
\end{equation}
and $n_\lambda (x)$ measures the multiples of $2\pi$ with which the
plaquette variable is shifted away from the interval $[\pi,\pi)$:
\begin{equation} 
        n_\lambda (x) = \frac{1}{2\pi} \epsilon_{\lambda\mu\nu}
        \left\{\theta_{\mu\nu}(x) - [\theta_{\mu\nu}(x)]_{2\pi}\right\}.
\end{equation}
Here, we use the usual modulo operation $[a]_{2\pi} \equiv \alpha - 2\pi n$ which subtracts $n$ multiples of $2\pi$ from the variable $\alpha$ such that $[a]_{2\pi}$ takes values in the interval $[\pi,\pi)$.
While $l_\mu(x)$ measures the quantized vorticity, $n_\lambda(x)$
gives the number of elementary Dirac strings piercing the plaquette
with its normal pointing in the $\lambda$ direction. 

Monopoles are detected by taking the divergence of $j_\mu$, $m(x) =
\nabla_\mu j_\mu(x)$, where $m(x)$ takes on integer values only.  Using
these definitions, we record the vortex line and monopole densities
\begin{equation} 
v \equiv \frac{1}{L^3}\sum_{x,\mu} \left|j_{\mu}(x)\right|, \quad M
\equiv \frac{1}{L^3} \sum_x m(x).
\end{equation} 
As already mentioned in the Introduction, in the $|\phi|^4$ theory, $v$
behaves similar to the energy and the associated susceptibility similar
to the specific heat \cite{bittner3}. A short summary of results for the
observables not involving vortices can be found in
Ref.~\cite{Wenzel:2005nd}.

\begin{figure}
\begin{minipage}{0.5\textwidth}
\includegraphics[width=0.8\textwidth]{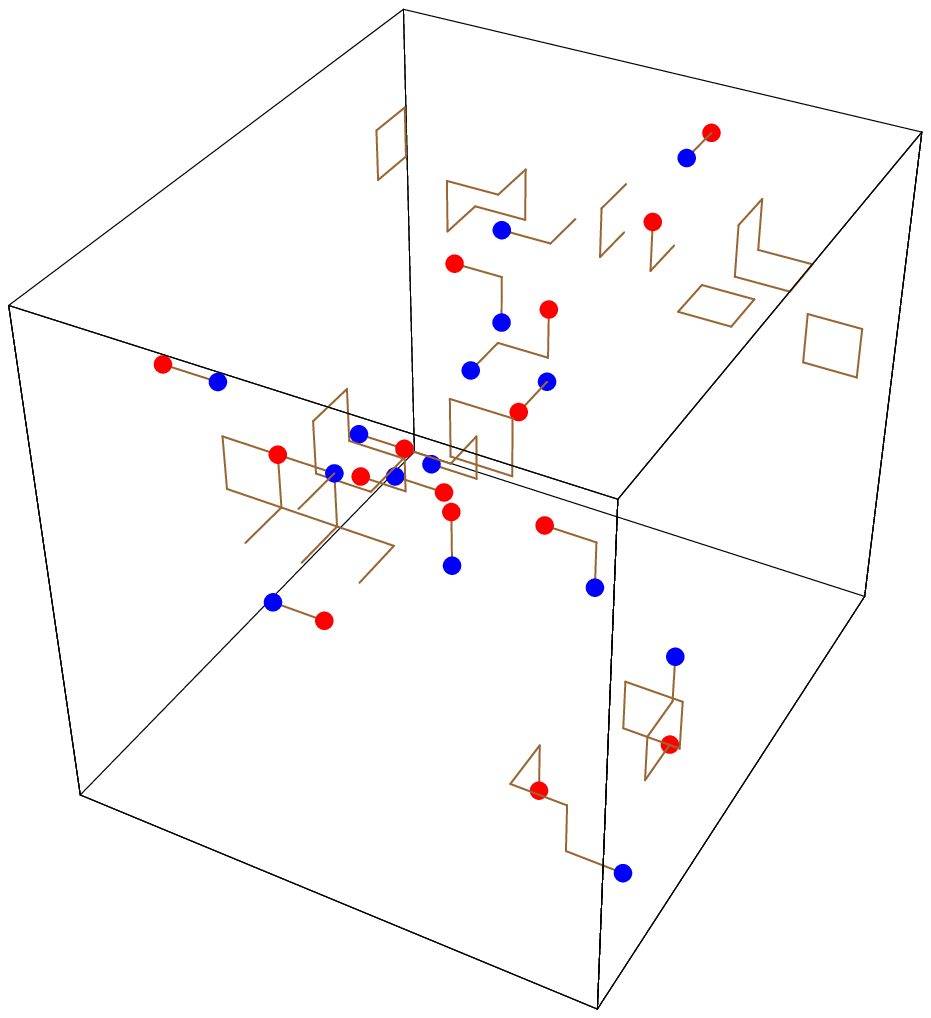}
\end{minipage}
\begin{minipage}{0.5\textwidth}
\includegraphics[width=0.8\textwidth]{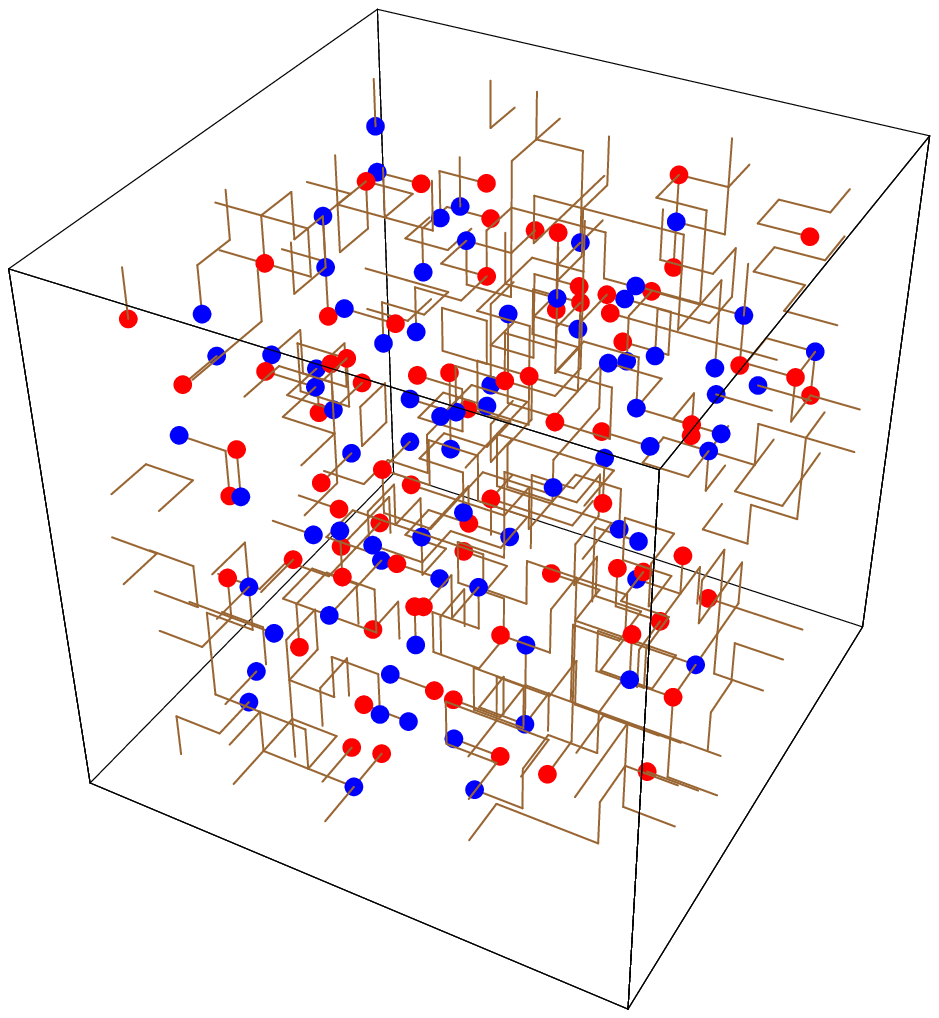}
\end{minipage}
\caption{\label{fig:networks_vis} Networks formed by vortex lines. The
  small spheres represent monopoles and antimonopoles. \textit{Left}:
  Higgs phase, characterized by a few short vortex lines which can
  either be closed or open.  \textit{Right}: Confinement phase,
  characterized by a large percolating vortex network.}
\end{figure}

The main focus in this paper is on the vortex networks formed by
individual vortex lines (see Fig.~\ref{fig:networks_vis}). Tracing out
such a network is ambiguous.  We restrict ourselves to the simplest
convention by defining a vortex line, which can be either open or
closed, as a set of connected vortex line segments.  Line segments are
said to be connected if they enter or leave the same lattice cube.  With
this convention, four or six line segments entering and leaving a single
cube are not further resolved into separate vortices, but are lumped
together into one, self-intersecting vortex. The size $n$ of a vortex is
just the number of links forming the vortex.  Vortex line segments with
$|j_\mu|\geq 2$ are not split into distinct segments and are only
counted once in the vortex density.  The vortex networks found with the
help of these tracing rules are therefore only an approximation to the
true networks.  And the observed, or apparent percolation threshold will
in general differ from the true one.
We trace out the vortex networks by using a recursive algorithm
and employ standard percolation observables to probe them
\cite{StaufferAharony}. Specifically, we determine the percolation
probability $P_\mathrm{1D}$ by recording when a vortex spans the lattice
in an arbitrary direction.  This definition is such that a vortex is
already said to percolate when it spans the lattice in just one,
arbitrary direction.  We also determine the probability $P_\mathrm{3D}$
that a vortex spans the lattice in all directions, and the percolation
strength $P_\infty$, defined as the size of the largest vortex per
lattice site.
We have in addition considered the average vortex size, but this data is
not well suited for analyses as we also found in previous studies on
simpler models \cite{janke-2006}.  Finally, we use the susceptibility
\begin{equation} 
\chi(\mathcal O)=\left(\langle
{\mathcal O}^2 \rangle - \langle {\mathcal O} \rangle^2\right)L^3\,,
\end{equation}  
(without the $L^3$ for $P_{1\mathrm{D}}$ and $P_{3\mathrm{D}}$) and the Binder parameter 
\begin{equation} 
  B(\mathcal{O})=1-\frac{\langle {\mathcal O}^4 \rangle}{3\langle {\mathcal O}^2
    \rangle^2}
\end{equation}  
of an observable $\mathcal{O}$ to probe the phase boundary.

\section{Simulation and Data Analysis Techniques}
\label{sec:Simulation}
\subsection{Monte Carlo }
A variety of Monte Carlo methods are applied to efficiently simulate the
system in the different parts of the phase diagram.  In the first-order
transition region, primarily the multicanonical algorithm (MUCA) \cite{MUCA} is used with a weight iteration as described in
Ref.~\cite{janke_histogramms}. Weights are rendered flat in the hopping
term (\ref{hop}) of the action to enhance tunnelling.  In addition to
local updates, the Higgs amplitude and the gauge angles are also updated
globally \cite{kajantie-1996-466} to allow for larger jumps in phase
space and thus for shorter tunnelling times between the two metastable
states.  In the vicinity of the Kert\'esz line, the gauge fields are
updated using the Metropolis algorithm, while the Higgs field is updated
by means of heat-bath and overrelexation algorithms
\cite{bunk:heatbath}.

Measurements are typically taken after each sweep of the lattice and the
entire time series is recorded to allow for error analysis and
post-simulation data processing. Table~\ref{tab:simparameters} provides
an overview of parameters used in the simulations.
\begin{table}[b!]
\caption{\label{tab:simparameters}Number of sweeps of the lattice of linear size $L$ and the
number of different values of $\kappa$ considered for each lattice used
in the simulations.}\vspace{0.2cm}\centering
\begin{tabular}{|c|c|c|c|c|}
\hline\hline
$\beta$ & $\lambda$ & $L$ & sweeps & different $\kappa$ values  \\
\hline
$1.1$ & $0.2$ & $10,12,14,16,18,20,22,24,26,30,32,42$ & $10^5$ -- $10^6$ & 20 \\
      & $0.75$ & $8,10,12,16,20$ & $10^5$ & 20 \\
      & $0.025$ & $8,10,12,14,16,18,20$ & $10^7$  & 1 (MUCA) \\
\hline
$2.0$ & $0.2$ & $8,10,12,14,16,20,24,28,36$ & $10^6$ & 10 \\
\hline\hline
\end{tabular}
\end{table}
An integral part of our analysis are Ferrenberg-Swendsen reweighting
techniques \cite{ferrenberg,ferrenberg_multi}. These techniques are
applied in two distinct flavours. In the Kert\'esz region, we use the
multihistogramm reweighting form, which means that we combine
simulations at different parameters in an optimal way.  Standard
reweighting with MUCA weights is applied in the first-order transition
region.  To apply these techniques in the best possible way we use
optimization routines such as the Brent method
\cite{Numerical-Recipes,brent} to search for maxima in susceptibilities
or crossing points of Binder parameters.  The use of the Jackknife
method \cite{efron} on top of these methods allows for error estimation
of the results so determined.

\begin{figure}[t!]
    \begin{center}
      \includegraphics[width=0.6\textwidth]{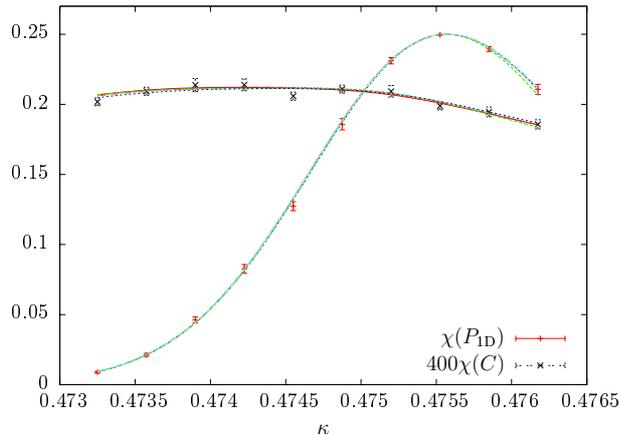}
      \caption{\label{fig:compare} Data points for the
        susceptibilities $\chi(P_\mathrm{1D})$ and $\chi(C)$
        (multiplied by $400$) for $\beta=2.0\,,\lambda=0.2$ and
        different values of $\kappa$ and their interpolating lines
        obtained through reweighting.  The measurements are taken on a
        cubic lattice of linear size $L=36$.}
\end{center}
  \end{figure}

As an example of this approach, we show in Fig.~\ref{fig:compare} the
susceptibilities of the percolation probability $P_\mathrm{1D}$ and
the coslink energy $C$. Note that the peak height of
$\chi(P_{1\mathrm{D}})$ is exactly $1/4$, since $\langle P_{1\mathrm{D}}^2 \rangle = \langle P_{1\mathrm{D}}\rangle$ and hence
$\chi(P_{1\mathrm{D}})=\langle P_{1\mathrm{D}} \rangle (1 - \langle
P_{1\mathrm{D}}\rangle)$ is maximal at $\langle P_{1\mathrm{D}}
\rangle =1/2$.  The points in the plot correspond to individual
simulations and the lines through the data points are obtained by
reweighting. Different lines in the figure correspond to having
different Jackknife blocks (but the same block in each time series)
omitted. Notice that the peak in the percolation susceptibility is
well defined whereas the peak in the coslink susceptibility is rather
broad.  The flatness of the peak is reflected in larger error bars on
the estimated peak location.

We have carefully checked our methods against known results for limiting
cases such as the London limit $\lambda \to \infty$, where fluctuations
in the Higgs amplitude are completely frozen, and the XY model, obtained
by setting in addition $\beta=\infty$.  A good check whether vortex
lines and monopoles are correctly identified is to perform a gauge
transformation under which their locations are invariant.

\subsection{Finite-size scaling methods}

We estimate critical exponents through finite-size scaling analyses.
The scaling Ansatz for an observable $\mathcal{O}$ states that in the
vicinity of a continuous phase transition, the data obtained for
different values of the tuning parameter $\kappa$ and for different
lattice sizes $L$ fall onto a (weakly) universal scaling function $f_\mathcal{O}$ defined through
\begin{equation} 
\label{unicurve}
\mathcal{O}_L(\kappa) = L^{\lambda_\mathcal{O}/\nu}
f_\mathcal{O}\left(t L^{1/\nu}\right), \quad t \equiv
(\kappa-\kappa_c)/\kappa_c .
\end{equation} 
Here, $\kappa_\mathrm{c}$ denotes the critical point, $\nu$ the correlation length exponent, and
$\lambda_\mathcal{O}$ is the critical exponent characterizing the
observable $\mathcal{O}$.  The variable $t$ is the reduced coupling and measures the distance from
the critical point.

Data collapse is usually a good check whether the right critical
exponents and critical couplings are found by other means.  With the
correct values, the measured data should fall on the universal curve
given by Eq.~\eqref{unicurve}.  Here, we reverse this idea. Starting
from an initial guess for the critical exponents and couplings, we compute the rescaled observable $\widehat{\mathcal{O}}_{L}(x)=\mathcal{O}_L(\kappa)L^{-\lambda_\mathcal{O}/\nu}$ with $x=tL^{1/\nu}$
and judge the quality of the collapse in the interval
$[x_\mathrm{min},x_\mathrm{max}]$ by introducing the weight function
  \begin{equation} 
    \sigma^2_\mathcal{O} \equiv \int_{x_\mathrm{min}}^{x_\mathrm{max}} \, 
    \mathrm{d}  x \left[\overline{\widehat{\mathcal{O}}_L^2}(x)  - 
      \overline{\widehat{\mathcal{O}}_L}^{\;2}(x)\right] ,
  \end{equation} 
  with $\overline{\widehat{\mathcal{O}}_L}(x) \equiv \sum_L \widehat{\mathcal{O}}_L(x)/
  n_L$ and $n_L$ the number of different lattice sizes included in the
  sum $\sum_L$. A perfect collapse in the window
  $[x_\mathrm{min},x_\mathrm{max}]$ would correspond to
  $\sigma_\mathcal{O}^2= 0$, whereas a bad collapse has a large
  $\sigma^2_\mathcal{O}$.  This approach qualifies for implementation as
  an automated method when combined with optimization algorithms to
  minimize $\sigma^2_\mathcal{O}$ by adjusting the values of the
  critical exponents.  The method requires that data in each point in
  the interval $[x_\mathrm{min},x_\mathrm{max}]$ of the rescaled $x$-axis
  be compared, most of which has not been measured.  A possibility
  is to interpolate between data points by using a polynomial expansion
  of the function $f_\mathcal{O}$ and to fit its coefficients to the
  data $\mathcal{O}_L$ \cite{sandvik:collaps2:2006}.  The fit then
  allows for calculating $\sigma_\mathcal{O}^2$.  We use a different
  approach in that we apply the standard reweighting techniques
  mentioned above to calculate $\mathcal{O}_L$ at every point $\kappa$ and hence  $\widehat{\mathcal{O}}_L$ at every point $x$. We feel that this approach is more natural and does not add further degrees of freedom to the process.

  In practice our implementation is as follows. We first reweight the
  measured data on all system sizes.  We then start with initial values
  for the exponents and couplings, and apply a minimization algorithm,
  such as the simplex method \cite{Numerical-Recipes,gsl}, that varies
  the exponents until $\sigma^2_\mathcal{O}$ is minimized. Since the
  algorithm can become locked in a local minimum, the procedure must be
  repeated for many starting points. We find that for our purposes,
  where only three parameters need to be determined, the routine gives
  reliable and consistent results. To obtain error estimates, we apply
  the Jackknife method on top of the whole process. The effect of
  correction terms to scaling is minimized by repeating the procedure
  for increasingly smaller intervals $[x_\mathrm{min},x_\mathrm{max}]$
  around the point where collapse is attempted. We expect our tool to be
  useful in other simulation studies as well.  An idea similar to our
  implementation was recently presented in
  Ref.~\cite{seno:collaps:2003}.

\section{First-Order Transition Region}
\label{sec:1st}
We start our numerical study by investigating the behavior of the vortex
network in the first-order transition region of the phase diagram sketched
in Fig.~\ref{fig:phase_diagram}.  We choose to simulate at $\beta=1.1$
and $\lambda=0.025$ where the first-order transition is strong enough
already on small lattices so that time consuming simulations on larger
systems are not needed.
\begin{figure}[b]
  \begin{minipage}{0.5\textwidth}
    \includegraphics[width=0.95\textwidth]{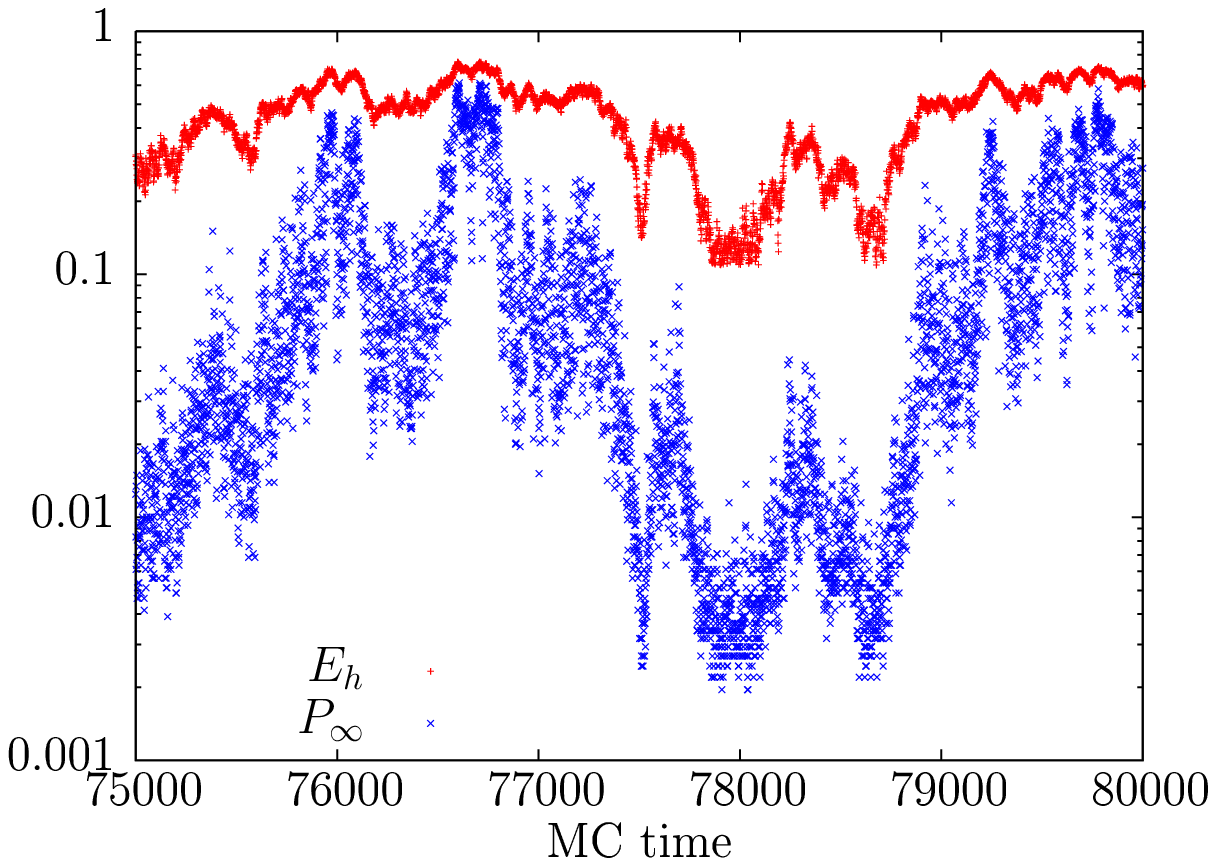}
  \end{minipage}
  \begin{minipage}{0.5\textwidth}
    \includegraphics[width=0.95\textwidth]{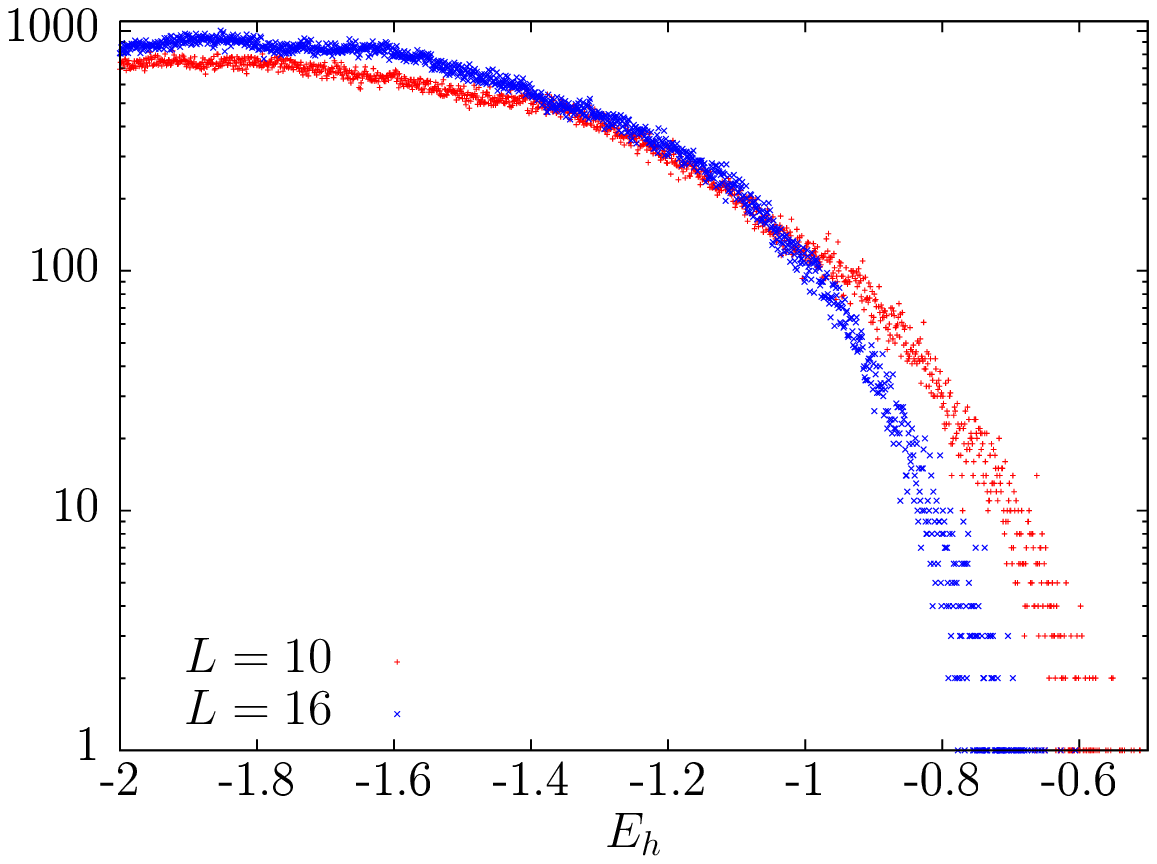}
  \end{minipage}
  \caption{\label{fig:timeseries}\textit{Left}: Time series of a MUCA
    simulation showing the hopping energy $E_\mathrm{h}$ (upper curve)
    and percolation strength $P_\infty$ (lower data points) for $L=16$.
    The two observables are seen to be intimately connected.  To fit in
    one plot, the hopping energy is shifted upwards.  \textit{Right}:
    Correlation histogram of the number of configurations \textit{without} a
    percolating network over hopping energy $E_\mathrm{h}$ for system sizes $L=10$ and $16$. Note the logarithmic scale used on the vertical axis.}
\end{figure}
The left plot in Fig.~\ref{fig:timeseries} shows time series of the
hopping energy $E_\mathrm{h}$ and the percolation strength
normalized by the volume as measured in a typical MUCA simulation
\cite{MUCA}.  Changes in the energy, which reflect the more or less
random walk through the metastable region, are seen to occur jointly
with changes in the percolation strength.  The right plot shows the
correlation histogram of the number of configurations without a
percolating network versus hopping energy $E_\mathrm{h}$.  Larger
negative energies are seen to strongly correlate with the absence of a
percolating vortex network while smaller negative energies strongly
correlate with the presence of such a network.

The central thesis we put forward in Ref.~\cite{Wenzel:2005nd} is that
the phase diagram of the model can be understood in terms of
proliferating vortices.  In the first-order region, the location of the
phase transition has been determined to high precision with the help of
observables not involving vortices such as the hopping energy.  To
determine the location of the percolation threshold in the
infinite-volume limit, we consider the vortex percolation probability
$P_\mathrm{1D}$ and percolation strength $P_\infty$, and analyze the
scaling of the locations $\kappa_\mathrm{per}(L)$ of the maxima in the
associated susceptibilities and Binder parameters with lattice size $L$.
At a first-order transition, $\kappa_\mathrm{per}(L)$ is expected to
scale as
\begin{equation}
\label{eqn:foscaling}
\kappa_\mathrm{per}(L)= \kappa_\mathrm{per} + c L^{-3} +
\mathcal{O}\left(L^{-6}\right),
\end{equation}
with $c$ a constant.  We reweight the MUCA time series to obtain the
susceptibilities and Binder parameters in the vicinity of
$\kappa_\mathrm{per}(L)$, and use the methods presented above to search
for the peak locations and its error bars.
\begin{figure}[t]
  \begin{minipage}{0.47\textwidth}
  \includegraphics[width=\textwidth]{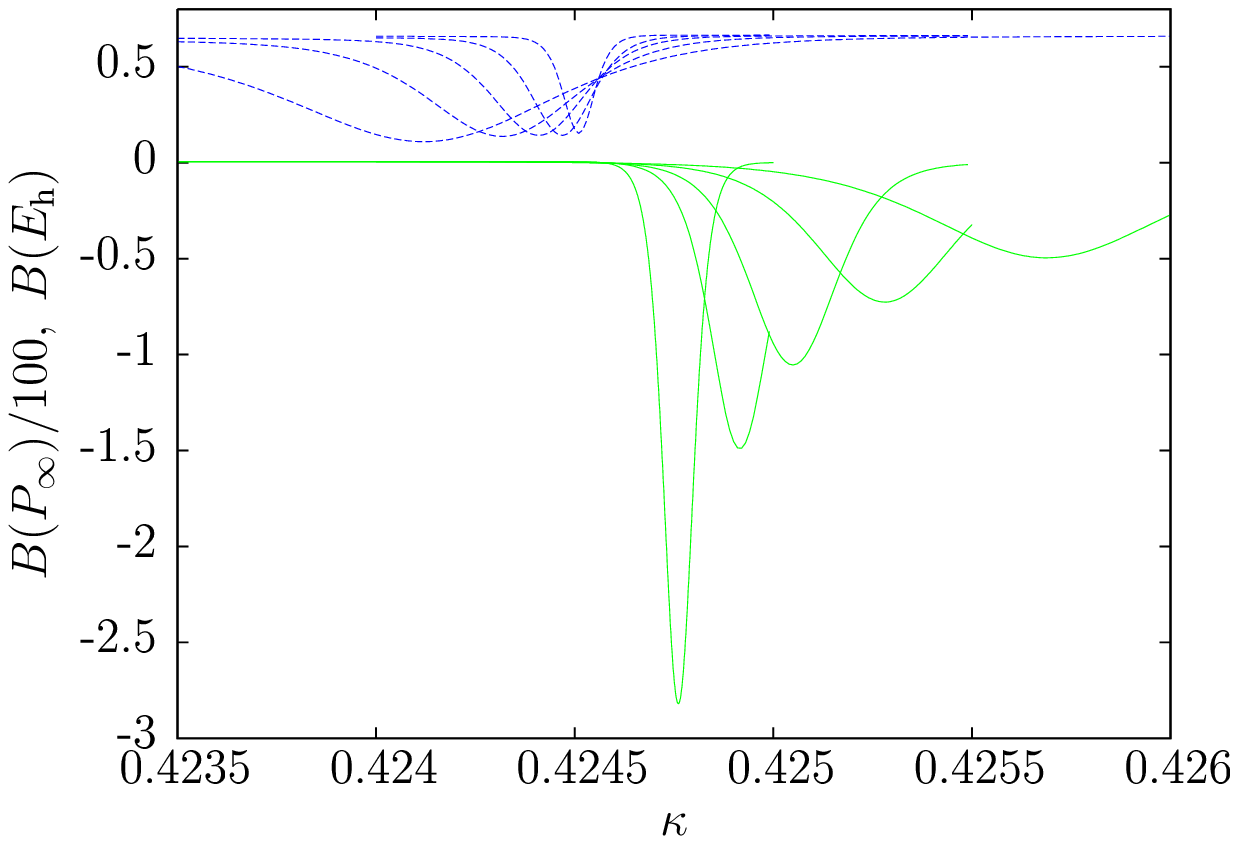}
  \end{minipage}
  \begin{minipage}{0.53\textwidth}
  \includegraphics[width=\textwidth]{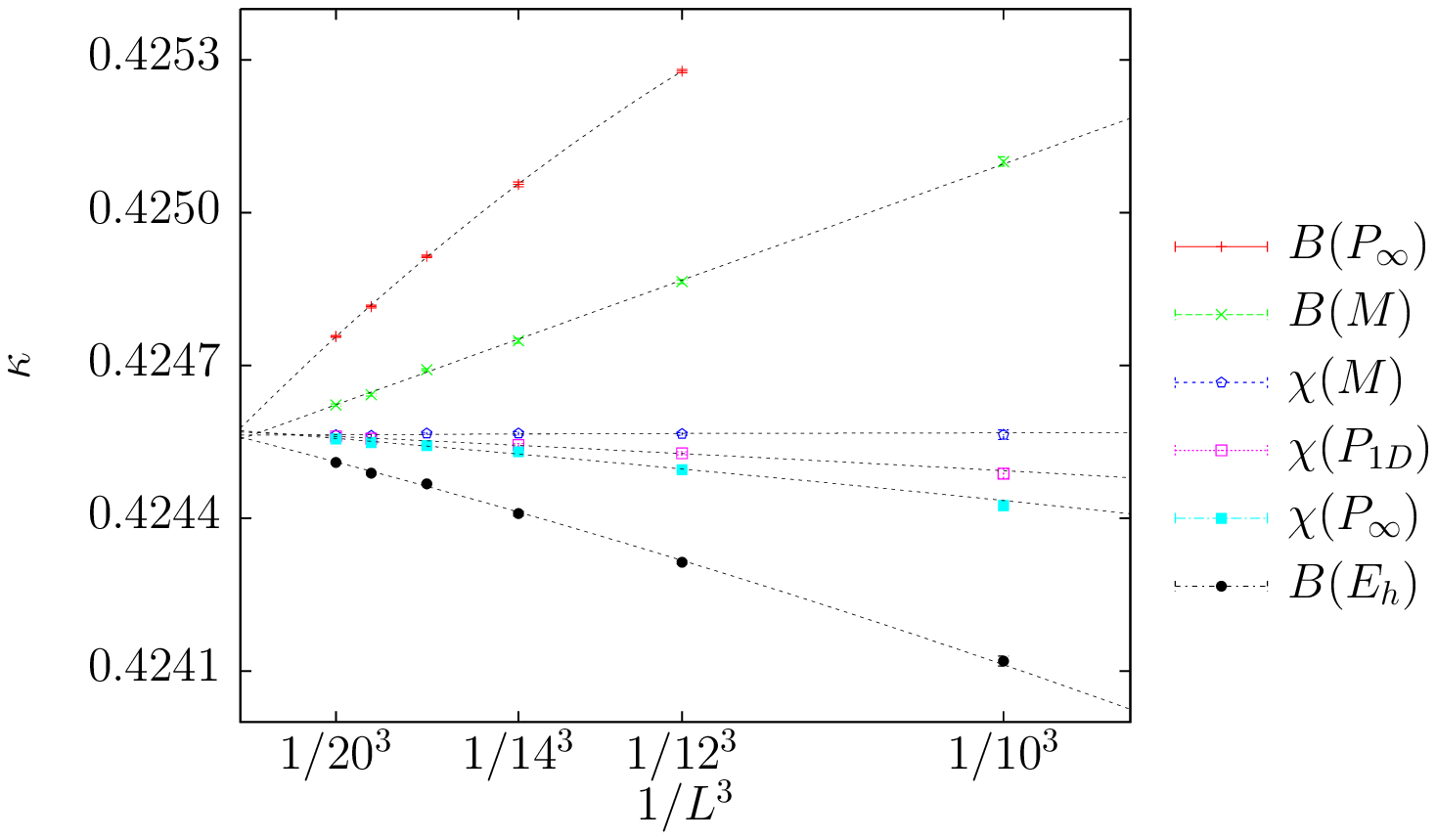}
  \end{minipage}
  \caption{\label{fig:scaling_fo} \textit{Left}: Binder parameters
    $B(E_\mathrm{h})$ (upper curves) and $B(P_\infty)$ (lower curves) as
    a function of $\kappa$ in the first-order transition region at
    $\beta=1.1$, $\lambda=0.025$ measured on lattices of linear size $L=
    10,12,14,16,20$ (from broad to sharp peaks). The Binder parameter
    $B(P_\infty)$ has been divided by 100 so that both parameters fit
    into one plot.  \textit{Right}: Scaling plots of the maxima of the
    various Binder parameters and susceptibilities. The Binder
    parameters show larger finite-size corrections.  All curves converge
    within error bars to the same infinite-volume value of
    $\kappa_\mathrm{per}=\kappa_\mathrm{c}=0.424570(3)$.  }
\end{figure}
Figure~\ref{fig:scaling_fo} summarizes our results. It shows the scaling
of the percolation thresholds $\kappa_\mathrm{per}(L)$ obtained from the
percolation probability $P_\mathrm{1D}$ and the percolation strength
$P_\infty$ together with the critical points $\kappa_\mathrm{c}(L)$
obtained from the hopping energy $E_\mathrm{h}$ and the monopole density
$M$.  The $\kappa_\mathrm{per}(L)$'s obtained from the network
observables are seen to strongly depend on the system size, while the
$\kappa_\mathrm{c}(L)$'s obtained from $\chi(M)$ are almost independent
of volume.  The results for $\chi(E_\mathrm{h})$ are not included in
Fig.~\ref{fig:scaling_fo} as they cannot be distinguished from those for
$M$ on the scale of the figure.  For lattice sizes larger than $L=12$,
all curves converge to the same point within error bars.  This brings us
to the important conclusion that the vortex percolation threshold is
located precisely at the first-order transition point. Our best estimate
for the transition point, based on both percolation and nonpercolation
observables, is $\kappa_\mathrm{c} = 0.424570(3)$ which is determined
from the average of the fit results summarized in
Table~\ref{tab:foresults}.  Averaging the estimates for percolation and
non-percolation observables separately, we obtain
$\kappa_\mathrm{per}=0.424572(4)$ and $\kappa_\mathrm{c}=0.424565(4)$,
respectively.  The absence of a mismatch between the critical
temperature and the vortex percolation threshold, as was found in the
$|\phi|^4$ theory using the same percolation observables
\cite{bittner3}, is probably because the phase transition is
discontinuous here.  An abrupt change in the ground state apparently
washes out any inaccuracy caused by an imperfect tracing out of the
vortex network.
\begin{table}[b]
  \caption{\label{tab:foresults}Results of fitting the data in
    Fig.~\ref{fig:scaling_fo} to Eq.~\eqref{eqn:foscaling}. The fits
    starting at $L = 8$ and $L =10$ also include the correction term
    $1/L^6$.  Those starting at $L = 12$ include only the leading term
    $1/L^3$, except for $B(P_\infty)$, where the correction term is
    still needed to obtain reasonable results. The number in square
    brackets denotes $\chi^2$ per degree of freedom (DOF),
    $\chi^2/\mathrm{DOF}$. Results are grouped according to percolation and non-percolation observables.}\vspace{0.2cm}\centering
\begin{tabular}{c|c|c|c}
\hline\hline
Observable & $L\geq8$ & $L\geq10$ & $L\geq12$  \\
\hline
$\chi(P_\mathrm{1D})$ & $0.424569(5)$ $[1.6]$ & $0.424569(5)$  $[2.0]$ & $0.424570(6)$ $[2.3]$ \\
$\chi(P_\infty)$ &  $0.424572(6)$ $[1.9]$ &  $0.424572(5)$ $[2.2]$ & $0.424573(6)$ $[2.5]$ \\ 
$B(P_\infty)$ & $0.424608(10)$  $[10.7]$ & $0.424608(10)$ $[13.1]$ &  $0.424577(6)$ $[1.1]$ \\
\hline $\chi(M)$ & $0.424567(3)$ $[2.4]$ & $0.424566(5)$ $[3.1]$ & $0.424567(6)$ $[4.2]$ \\
$\chi(E_\mathrm{h})$ & $0.424564(5)$ $[2.52]$ & $0.424564(6)$ $[3.1]$ & $0.424564(6)$ $[3.2]$ \\
$B(M)$ &  $0.424558(5)$ $[2.6]$ & $0.424558(8)$  $[3.5]$ & $0.424555(5)$ $[3.2]$  \\
$B(E_\mathrm{h})$ & $0.424565(6)$ $[3.3]$ & $0.424568(5)$ $[3.9]$ & $0.424566(6)$ $[4.0]$ \\
\hline\hline
\end{tabular}
\end{table}
\begin{figure}[t]
\begin{minipage}{0.5\textwidth}
\includegraphics[clip, width=0.9\textwidth]{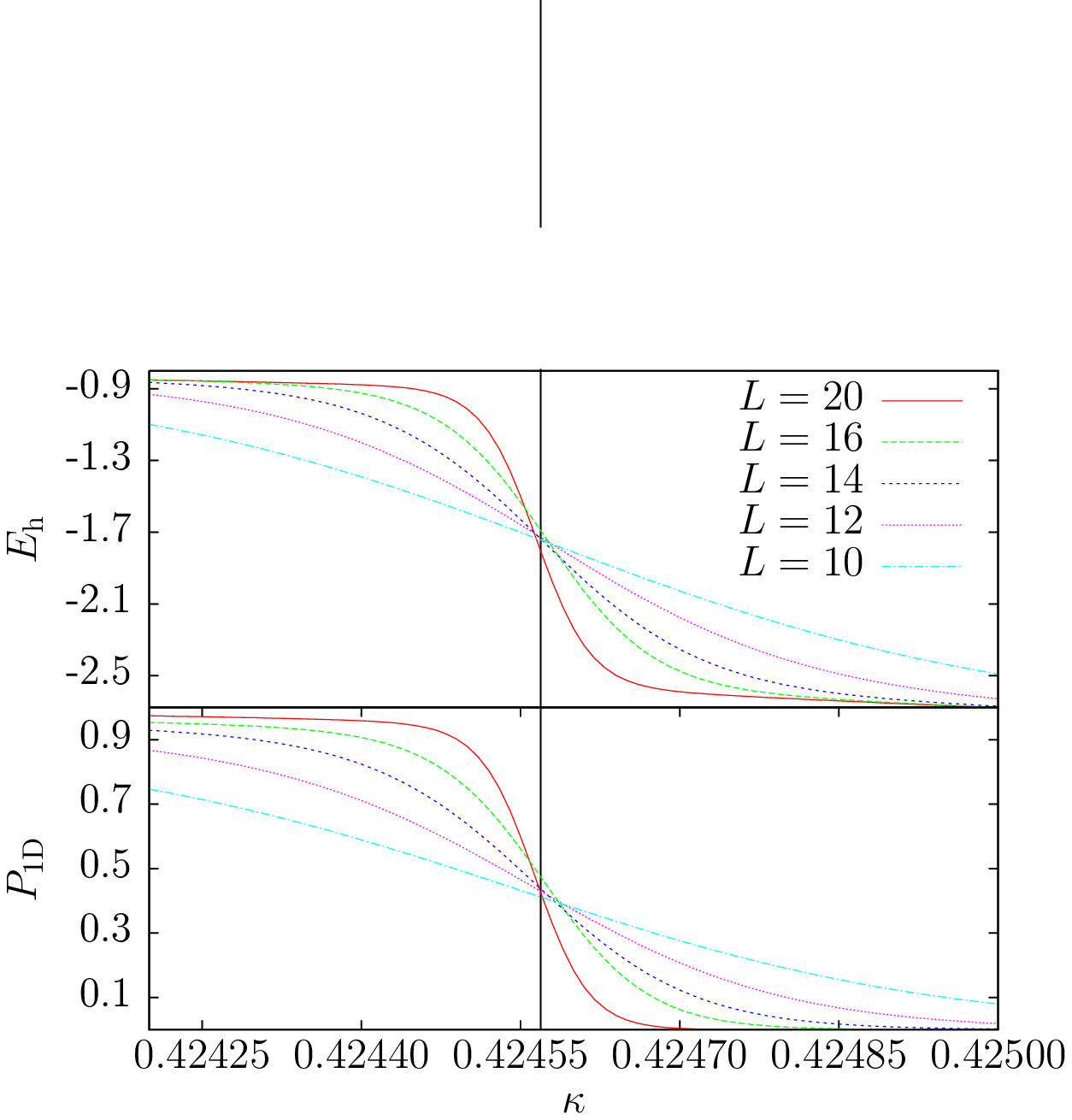}
\end{minipage}
\begin{minipage}{0.5\textwidth}
\includegraphics[width=0.9\textwidth]{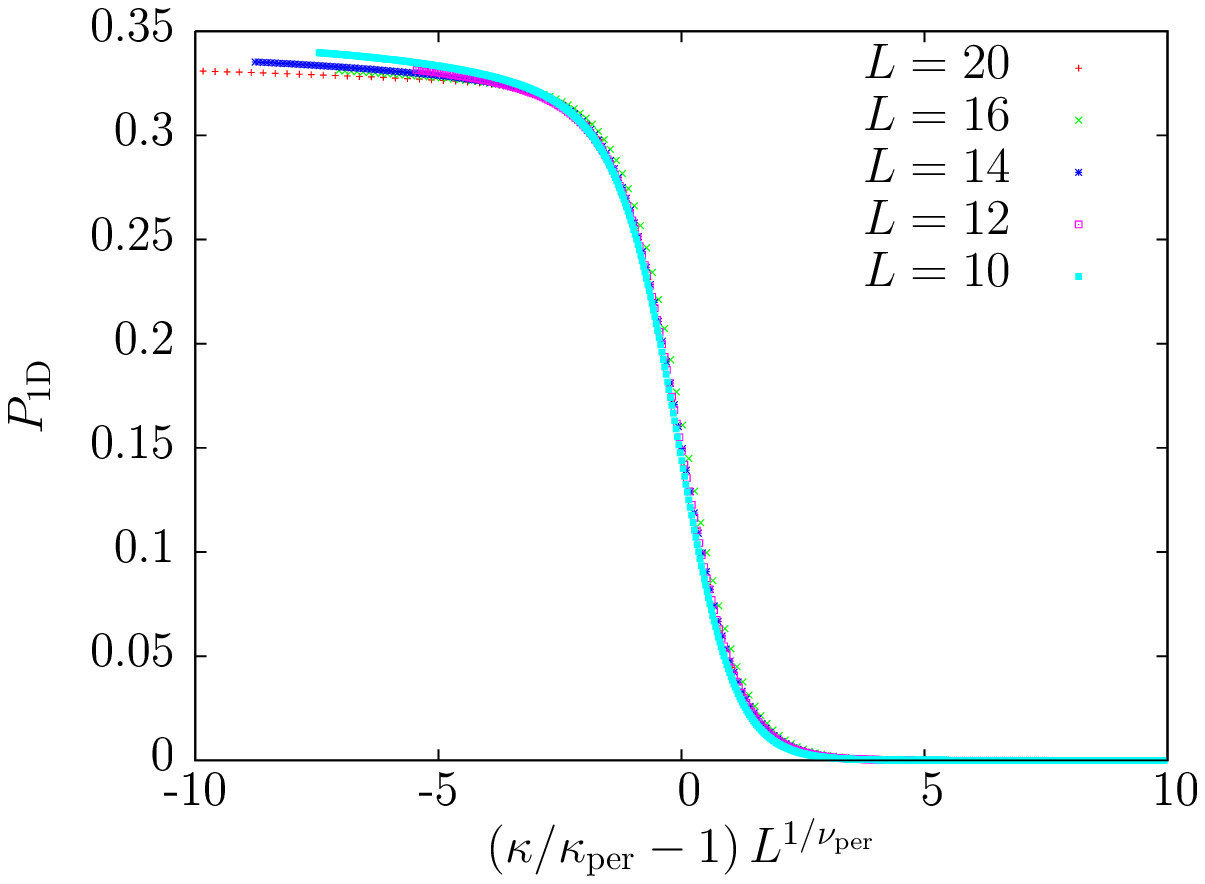}
\end{minipage}
\caption{\label{fig:fo_collaps1} \textit{Left}: Reweighted hopping
  energy $E_\mathrm{h}$ and percolation probability $P_\mathrm{1D}$ for
  various lattice sizes as a function of the hopping parameter close to
  the first-order phase transition.  The crossing points should provide
  a second method to estimate the transition point.  The estimates
  obtained using the two observables agree within error bars.
  \textit{Right}: Collapse of the $P_\mathrm{1D}$ data with
  $\kappa_\mathrm{per} = 0.424568, \nu_\mathrm{per} = 0.325$. An optimal
  collapse of the $E_\mathrm{h}$ data (not shown) is achieved for
  $\kappa_\mathrm{c} = 0.424563$ and $\nu=0.338$.}
\end{figure}

To establish in an unbiased fashion that the vortex network reflects the
first-order nature of the transition at $\beta=1.1$ and $\lambda=0.025$,
we assume nothing about the transition and apply standard finite-size
scaling to the vortex percolation probability $P_\mathrm{1D}$ as for a
continuous transition.  According to Eq.~(\ref{eqn:foscaling}), we then
expect to find $\nu_\mathrm{per}=1/d=1/3$.  To obtain an estimate for
the percolation threshold $\kappa_\mathrm{per}$ on the infinite lattice
and $\nu_\mathrm{per}$, we reweight the data points from the MUCA time
series and subsequently apply our collapsing routine with
$\kappa_\mathrm{per}$ and $\nu_\mathrm{per}$ as free parameters.
Figure~\ref{fig:fo_collaps1} shows the input and the result of this
procedure.  The extracted values, $\kappa_\mathrm{per} = 0.424568(6),
\nu_\mathrm{per} = 0.325(10)$, are perfectly consistent with a
first-order transition precisely at the expected location.  For
comparison, Fig.~\ref{fig:fo_collaps1} also shows the hopping energy
which is seen to display the same behavior as the percolation
probability.  Repeating this procedure for the hopping energy data, we
obtain $\kappa_\mathrm{c} = 0.424563(6)$ and $\nu = 0.338(10)$. Both
these estimates based on the data collapse analysis are perfectly consistent
with the previous estimates from finite-size scaling.  The raw data in
Fig.~\ref{fig:fo_collaps1} suggest as in \cite{PhysRevB.47.14757} that
the crossing points of the curves measured on lattices of different size
provide a second method to estimate the transition point.

\section{Kert\'esz Line}
\label{sec:kertesz}
We continue our analysis of the vortex network in the region where the
transition ceases to be of first order.  In Ref.~\cite{Wenzel:2005nd}, we
postulated that in this part of the phase diagram the Higgs and
confinement phases are separated by a Kert\'esz line.  Along this line
vortices proliferate, yet thermodynamic quantities remain nonsingular
across it.  The conjecture is based on the numerical observation that in
the Higgs phase, the monopoles are tightly bound in
monopole-antimonopole pairs. The magnetic flux emanating from a monopole
is squeezed into a magnetic flux tube (vortex) which ends on an
antimonopole.  The finite line tension forces the vortex lines to be
short.  In the confinement phase, the monopoles are no longer bound in
pairs, but form a plasma.  For this to arise, the vortex line tension
must vanish.  Vortex lines, both open and closed, can then grow
arbitrarily long at no energy cost and proliferate.
\begin{figure}[b]
 \begin{minipage}{0.5\textwidth}
   \includegraphics[width=\textwidth]{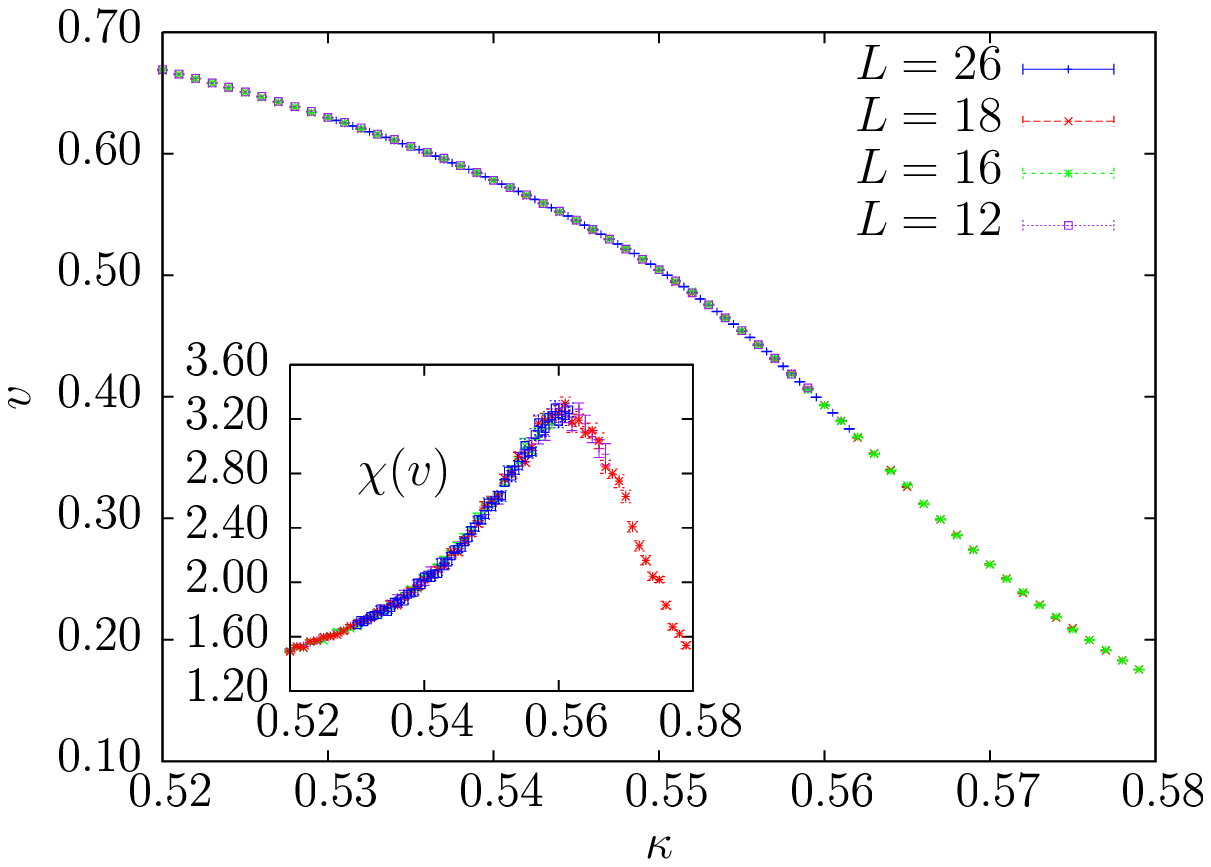}
  \end{minipage}
  \begin{minipage}{0.5\textwidth}
   \includegraphics[width=\textwidth]{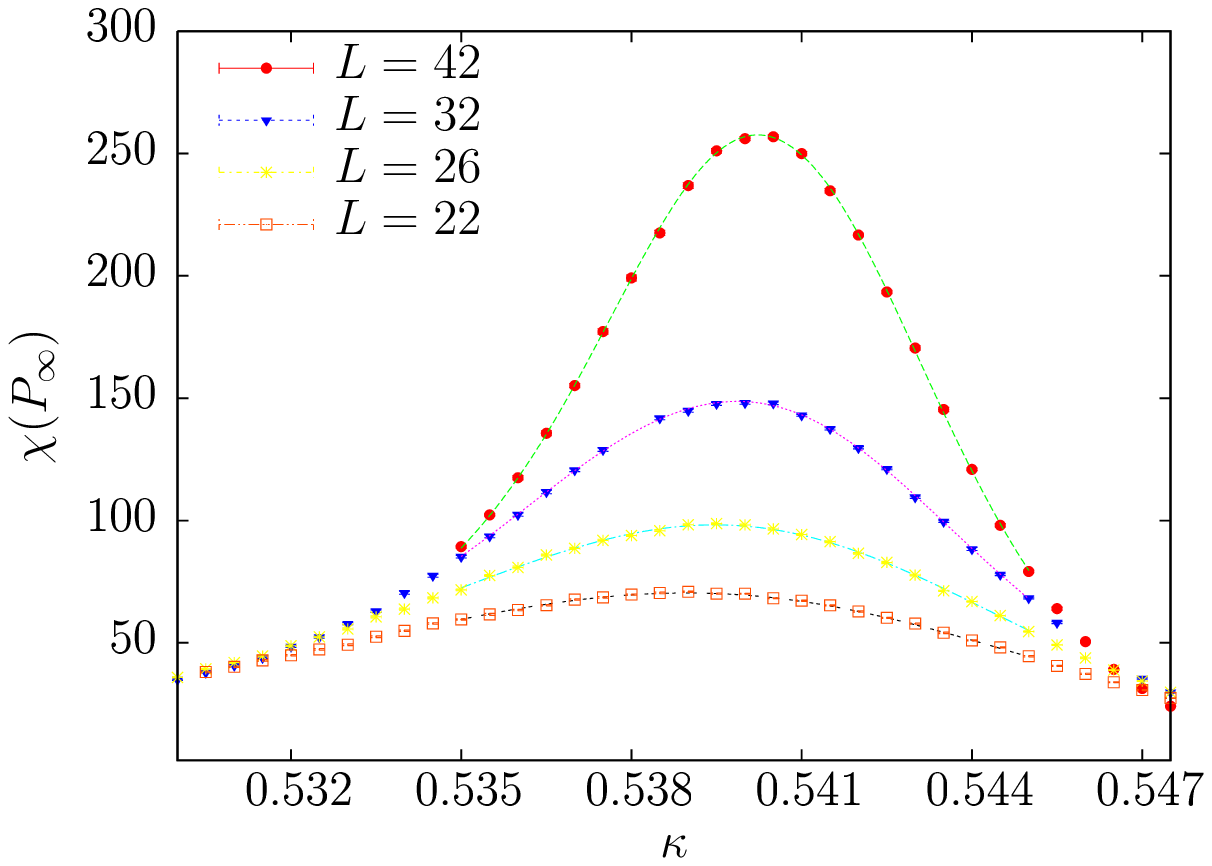}
  \end{minipage}
\caption{\label{fig:linedensity}\label{fig:chiPinfty} {\it Left:} Vortex line density and susceptibility (inset) as a function
      of the hopping parameter $\kappa$ for systems of linear size
      $L=12, 16, 18, 26$, and $\beta=1.1\,,\lambda=0.2$. {\it Right:} The susceptibility $\chi(P_\infty)$ as a function of the 
 hopping parameter $\kappa$ for systems of linear size $L=22, 26, 32, 42$.} 
\end{figure}
To facilitate comparison with our previous work \cite{Wenzel:2005nd}, we
choose the parameters $\beta=1.1$ and $\lambda=0.2$.
Figure~\ref{fig:linedensity} (left) shows our results for the vortex
line density as a function of the hopping parameter $\kappa$ for
lattices of linear size varying from $L=12$ to $L=26$.  The inset gives
the corresponding susceptibilities, which display the same remarkable
behavior first observed in the London limit $\lambda \to \infty$ for
other observables \cite{Bhanot:1981ug,schiller1}.  Namely, the
susceptibility data obtained on lattices of different sizes is seen to
collapse without rescaling.  In particular, the maxima of the
susceptibilities do not show any finite-size scaling.  From these data
we obtain the estimate $\kappa_\times= 0.5615(7)$ for the location of
the phase boundary.  Applying the same analysis to the hopping energy, we
arrive at the estimate $\kappa_\times= 0.5655(7)$ which is close, but
does not agree within error bars with the previous estimate.  To further
investigate this issue, we consider the monopole density to see if it
leads to the same estimate as does the hopping energy.  The results of
our initial study \cite{Wenzel:2005nd} indicated that both estimates
agree within error bars.  However, the high-precision Monte Carlo
simulations carried out in the present work show that this is not the
case for the monopole density yields $\kappa_\times=0.5639(1)$.  In
evaluating these results, it should be kept in mind that none of these
susceptibilities diverge in the infinite-volume limit, so that small
discrepancies were to be expected. The change in the ground state is for
this reason usually referred to as a crossover ($\times$).  The
discrepancy between the peak locations of these observables is shown
below in Fig.~\ref{fig:scaling_sus1}.

In Ref.~\cite{Wenzel:2005nd}, we argued that the phase diagram is more
refined than just showing a crossover between the Higgs and confinement
ground states in that a sharp boundary between the two phases does exist
in the form of a Kert\'esz line across which the vortices proliferate.
Moreover, as we argued partly on the basis of symmetry, the percolation
observables in the vicinity of the Kert\'esz line should be
characterized by the usual percolation exponents.  Unlike the
observables previously studied, these observables are expected to show
finite-size scaling.  Figure~\ref{fig:chiPinfty} (right) shows the
susceptibility $\chi(P_\infty)$ of the percolation strength as a
function of the hopping parameter $\kappa$ for lattices of linear size
varying from $L=22$ to $L=42$.  It is indeed observed that this
percolation observable depends on the lattice size even though we
considered system sizes larger than those for which the other observables
already reached the infinite-volume limit.  That is, in contrast to the
other observables, percolation observables allow for a precise location
of the phase boundary.

\begin{figure}[b]
 \begin{minipage}{0.5\textwidth}
    \includegraphics[width=0.95\textwidth]{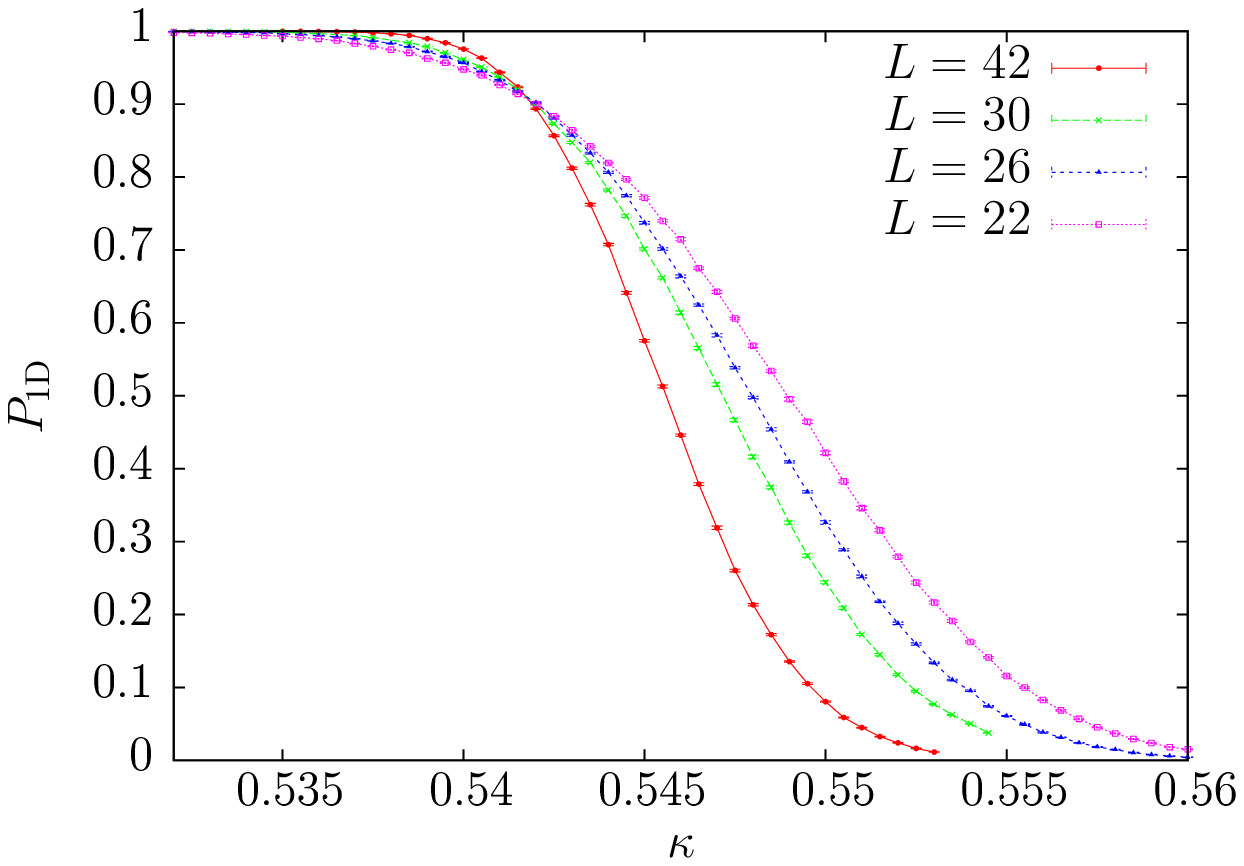}
  \end{minipage}
  \begin{minipage}{0.5\textwidth}
    \includegraphics[width=0.95\textwidth]{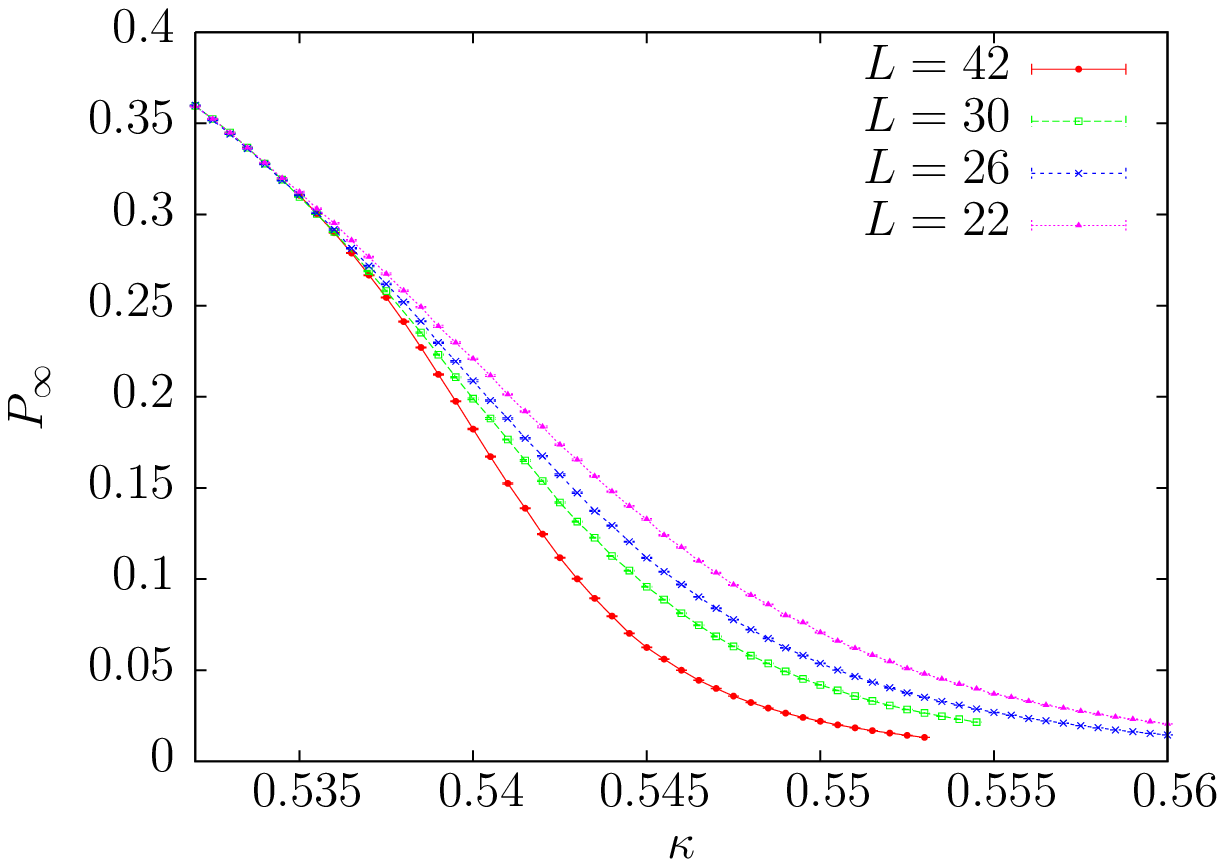}
  \end{minipage}
  \caption{Percolation probability $P_\mathrm{1D}$ and
    percolation strength $P_\infty$ as a function of the hopping parameter
    $\kappa$ for systems of linear size $L=22, 26, 30, 42$ at $\beta=1.1,\,\lambda=0.2$.
    \label{fig:Pchi}}
\end{figure}

To estimate the percolation exponents, we study the behavior of the
percolation probability $P_\mathrm{1D}$ and $P_\mathrm{3D}$ as well as
the percolation strength $P_\infty$ in the vicinity of the percolation
threshold $\kappa_\mathrm{per}$ (see Fig.~\ref{fig:Pchi}).  On the
infinite lattice, the percolation strength vanishes on approaching the
threshold as $P_\infty \sim (\kappa -
\kappa_\mathrm{per})^{\beta_\mathrm{per}}$, while $\chi(P_\infty)$
diverges as $\chi(P_\infty) \sim |\kappa -
\kappa_\mathrm{per}|^{-\gamma_\mathrm{per}}$.  Finally, the correlation
length $\xi_\mathrm{per}$, which provides a typical length scale of the
vortex network, diverges as $\xi_\mathrm{per} \sim |\kappa -
\kappa_\mathrm{per}|^{-\nu_\mathrm{per}}$.
Given the discrepancy found in the context of the $|\phi|^4$ theory,
we do not expect the estimate of the percolation threshold using
percolation observables to coincide with the one based on the vortex
line density.  We therefore determine the location of the percolation
threshold anew together with the exponent $\nu_\mathrm{per}$ by studying the
finite-size behavior of the susceptibilities of percolation
observables.  Figure~\ref{fig:scaling_sus1} shows the scaling of the
locations of the susceptibility maxima with $1/L$.  The observables
considered are the percolation probabilities $P_\mathrm{1D}$ and
$P_\mathrm{3D}$, and the percolation strength $P_\infty$. For
comparison, also the data for the hopping energy, which was used
in Ref.~\cite{Wenzel:2005nd} to estimate the location of the phase
boundary, and the data for the monopole and vortex densities are
included. 
\begin{figure}
    \begin{center}
      \includegraphics[width=0.8\textwidth]{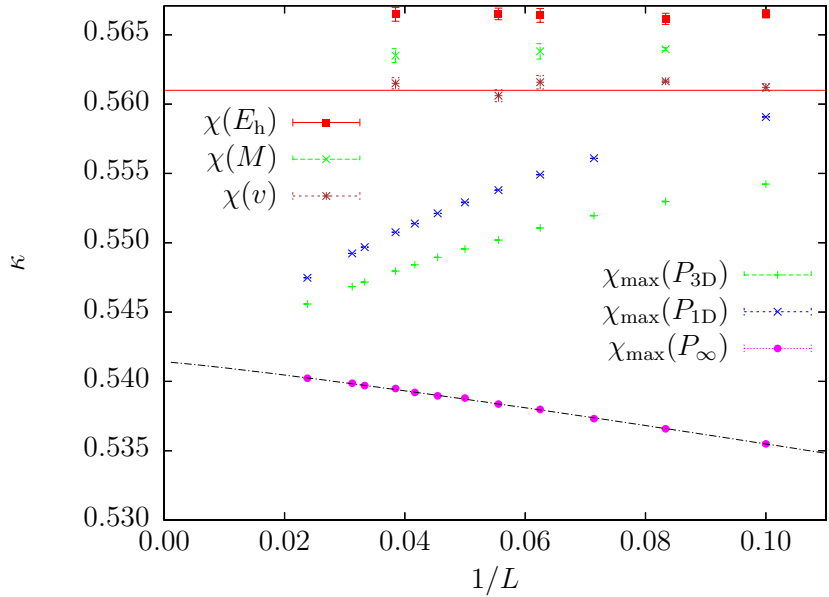}
      \caption{\label{fig:scaling_sus1} Locations of the susceptibility
        maxima of the percolation probabilities $P_\mathrm{1D}$ and
        $P_\mathrm{3D}$, and the percolation strength $P_\infty$ as a
        function of $1/L$ for $\beta=1.1$.  The dashed line is
        a fit of $\chi_\mathrm{max}(P_\infty)$ to
        Eq.~(\ref{eqn:peakscaling}).  The horizontal line is through the
        corresponding data for the vortex density, while the two sets of
        data points above this line pertain to the monopole density and
        hopping energy.}
\end{center}
  \end{figure}
The percolation data in Fig.~\ref{fig:scaling_sus1} are fitted to the
function
\begin{equation}
\label{eqn:peakscaling}
  \kappa_\mathrm{per}(L)=\kappa_\mathrm{per}+ c L^{-1/\nu_\mathrm{per}},
\end{equation}
using the standard least-squares method.  Unfortunately, this approach
does not give reliable and consistent values for $\kappa_\mathrm{per}$
and $\nu_\mathrm{per}$ when repeated for the different observables and
lattice sizes. The most stable results are obtained from the
susceptibility of $P_\infty$, giving $\kappa_\mathrm{per} =
0.54141(18)$ and $\nu_\mathrm{per} = 0.87(6)$. Results for
$P_\mathrm{1D}$ and $P_\mathrm{3D}$ depend too much on the fitting
regime and we conclude that Eq.~\eqref{eqn:peakscaling} for these
observables is only fulfilled for large lattice sizes. As expected,
the estimate of the location of the percolation threshold does not
agree with $\kappa_\times$ obtained from the vortex line density.
\begin{figure}[h!]
  \begin{minipage}{0.5\textwidth}
    \includegraphics[width=0.95\textwidth]{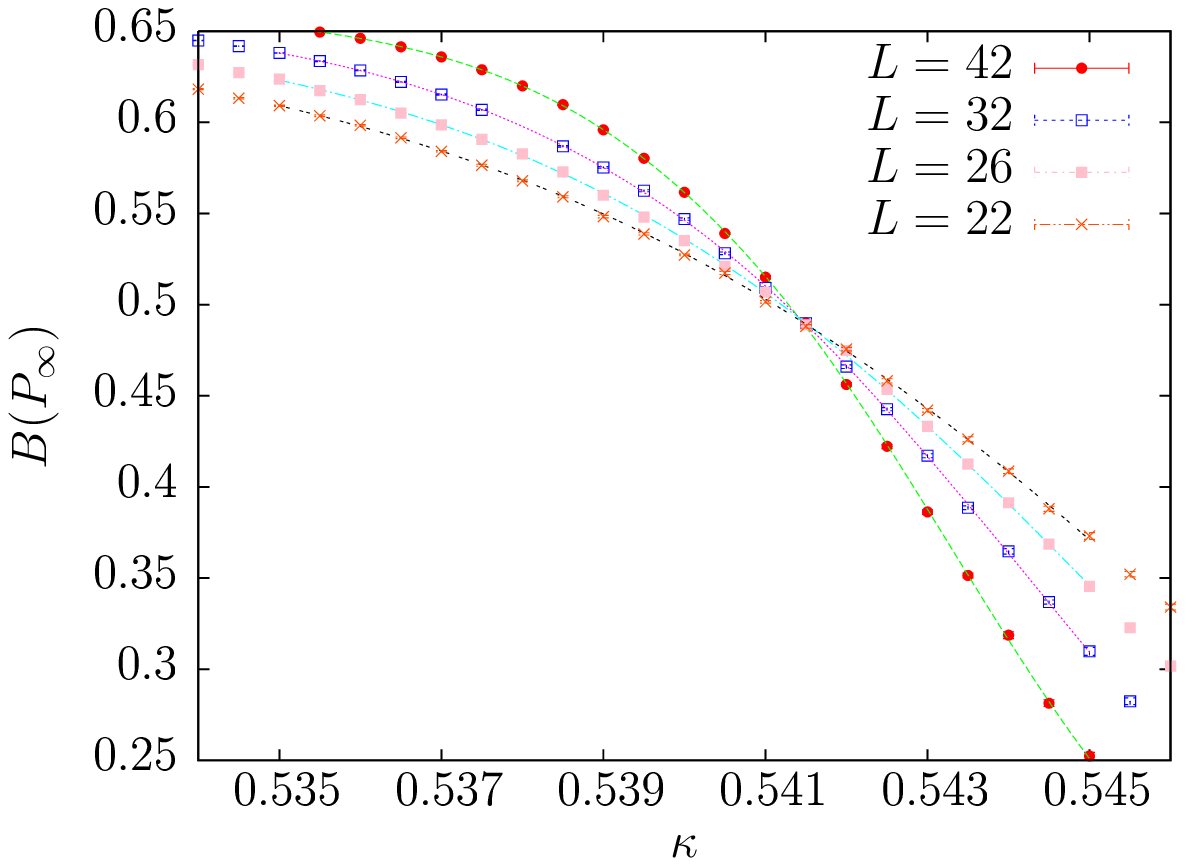}
  \end{minipage}
  \begin{minipage}{0.5\textwidth}
    \includegraphics[width=0.95\textwidth]{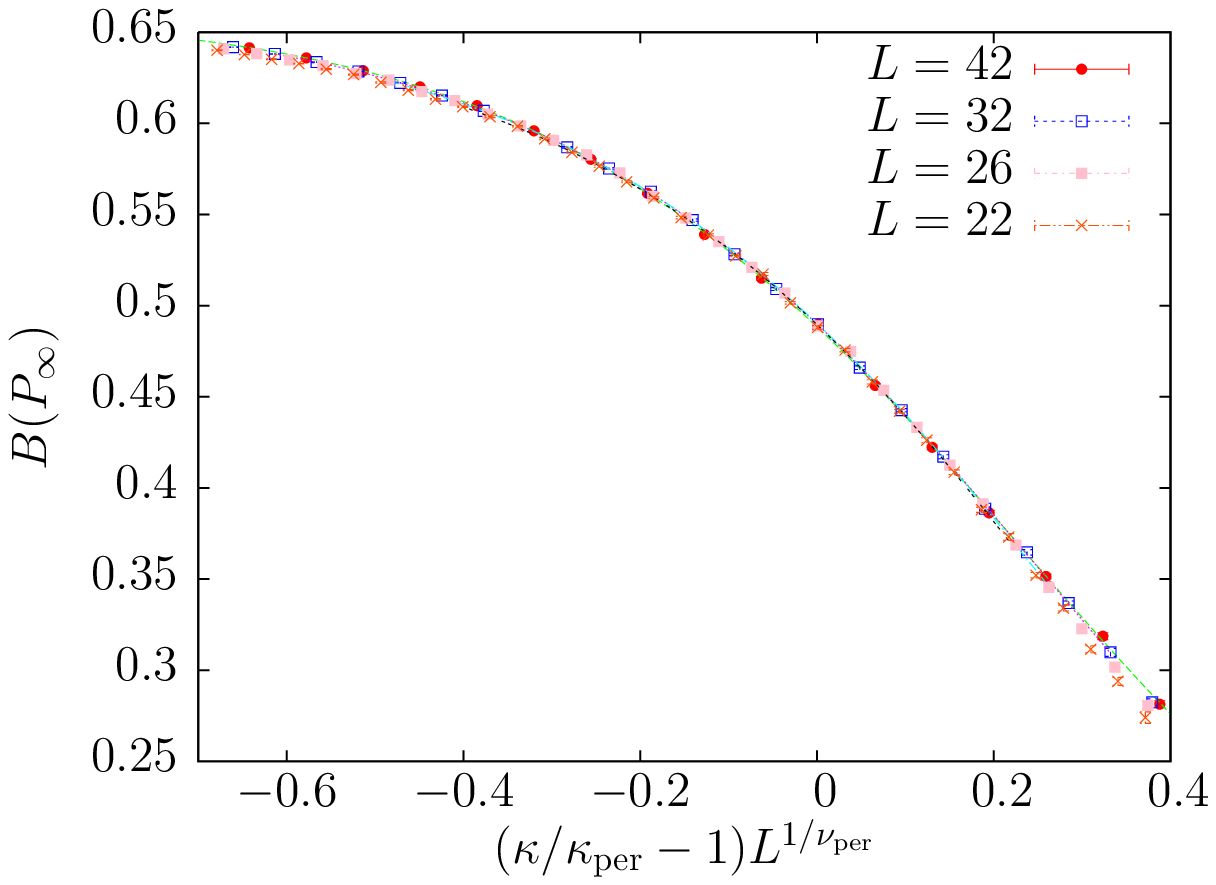}
  \end{minipage}
  \begin{minipage}{0.5\textwidth}
    \includegraphics[width=0.95\textwidth]{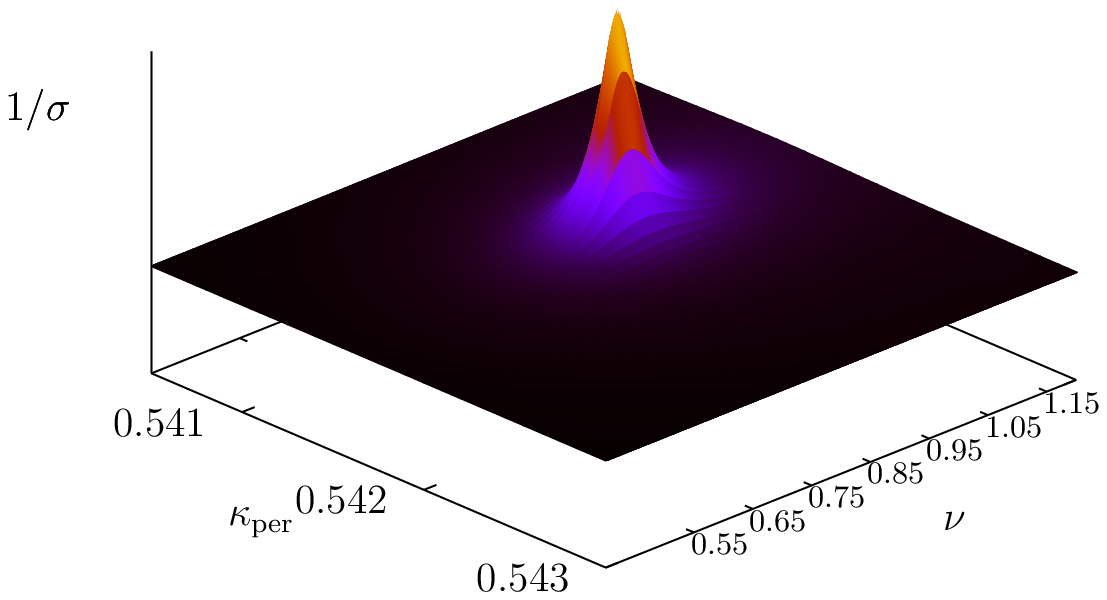}  
  \end{minipage}
  \begin{minipage}{0.5\textwidth}
  \caption{\label{fig:collaps_ANMAXSbinder1} \textit{Top Left}: Binder
    parameter of the percolation strength $P_\infty$ as a function of
    the hopping parameter $\kappa$.  \textit{Top Right}: Collapse of the
    data for $\kappa_\mathrm{per}=0.54145$ and $\nu_\mathrm{per}=0.881$.
    The data is obtained on lattices of linear size $L=22, 26, 30, 42$.
    {\it Bottom Left:} Landscape of the quality of the data collapse
    over a range of values of $\kappa_\mathrm{per}$ and
    $\nu_\mathrm{per}$. The peak corresponds to the best collapse and
    yields our estimates $\kappa_\mathrm{per}=0.54145(2)$ and
    $\nu_\mathrm{per}=0.881(2)$.}
   \end{minipage}
\end{figure}

\begin{figure}[t!]
  \begin{minipage}{0.5\textwidth}
    \includegraphics[width=0.95\textwidth]{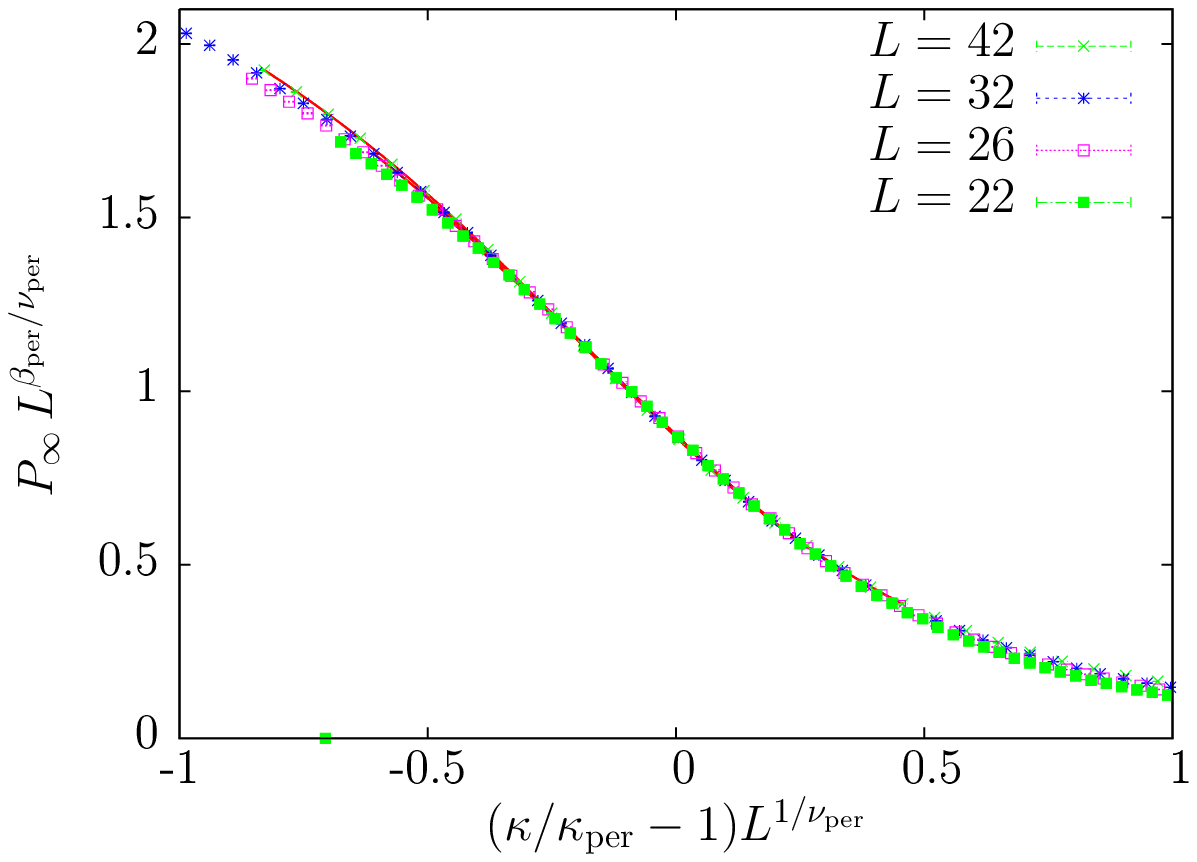}
  \end{minipage}
  \begin{minipage}{0.5\textwidth}
    \includegraphics[width=0.95\textwidth]{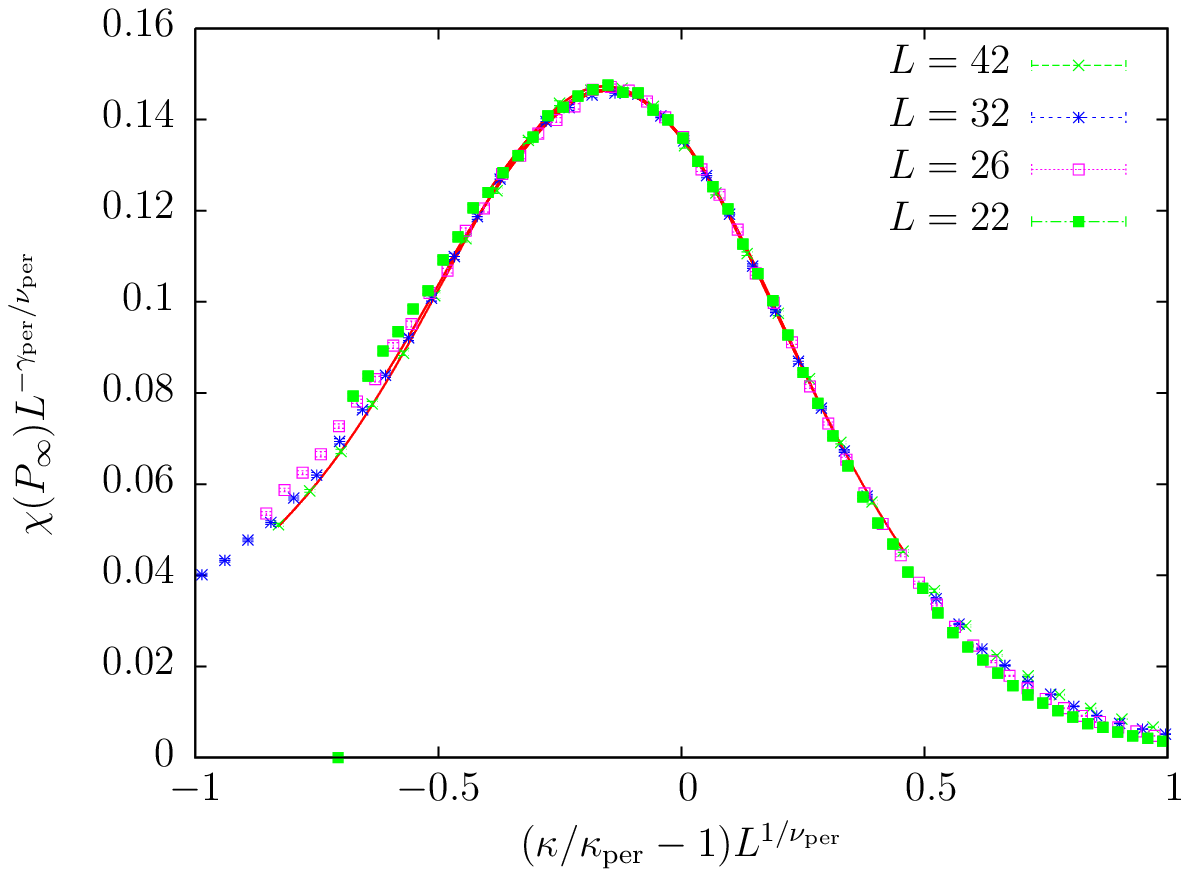}
  \end{minipage}
  \caption{\label{fig:collaps_ANMAXS1} {\it Left:} Collapse of the percolation strength $P_\infty$ yielding an exponent $\beta_\mathrm{per}=0.43(2)$. {\it Right:} The same for the susceptibility $\chi(P_\infty)$ yielding $\gamma_\mathrm{per}=1.76(2)$.
 }
\end{figure}

To improve these estimates, we now apply our collapsing routine, rather than
carrying out additional simulations for other lattice sizes. The method
has the advantage that much more data is used as input for not only the
locations of the peak maxima but all the data in the vicinity as well as
the interpolated values obtained from reweighting are included. We take
the Binder parameter $B(P_\infty)$ of the percolation strength together
with the scaling Ansatz
\begin{equation}
  B(P_\infty)=h\left(t L^{1/\nu_\mathrm{per}}\right) , \quad t \equiv
(\kappa-\kappa_\mathrm{per})/\kappa_\mathrm{per} ,
\end{equation}
where $h$ is a scaling function.
Figure~\ref{fig:collaps_ANMAXSbinder1} shows the raw data together
with the data collapse for $\kappa_\mathrm{per}=0.54145$ and
$\nu_\mathrm{per}=0.881$. The good quality of the data collapse is
apparent from the 3D plot in the same figure.
Given these estimates for $\kappa_\mathrm{per}$ and $\nu_\mathrm{per}$,
we next apply the collapsing routine to determine the exponents
$\beta_\mathrm{per}$ and $\gamma_\mathrm{per}$ from the scaling
relations
  \begin{equation}
    P_\infty=L^{-\beta_\mathrm{per}/\nu_\mathrm{per}}f\left(t
    L^{1/\nu_\mathrm{per}}\right) , \quad \chi(P_\infty) =
    L^{\gamma_\mathrm{per}/\nu_\mathrm{per}}
    g\left(t L^{1/\nu_\mathrm{per}}\right) 
\end{equation}
with $f$ and $g$ scaling functions.  Table~\ref{table:ces} summarizes
our results and compares them with the random percolation exponents.
\begin{table}[b]
\caption{\label{table:ces}Critical exponents of the percolating vortex 
  network across the Kert\'esz line for $\beta=1.1$ and $\lambda = 0.2$ 
  compared with the standard percolation exponents.}
\begin{center}
  \begin{tabular}{|c|c|c|c|c|}
    \hline\hline Model  &
    $\kappa_\mathrm{per}$ & $\nu_\mathrm{per}$ & $\beta_\mathrm{per}$ &
    $\gamma_\mathrm{per}$ \\ \hline cAHM & 0.54145(2) & 0.881(2) &
    0.43(2) & 1.76(2) \\ \hline Percolation \cite{ballesteros-1999-32} (see also \cite{deng:016126,hellmund:051113})     
    & - & 0.8765(16) & 0.4522(8) & 1.7933(85) \\ \hline\hline
  \end{tabular}
\end{center}
\end{table}
In agreement with our conjecture \cite{Wenzel:2005nd}, the estimates,
which as shown in Fig.~\ref{fig:collaps_ANMAXS1} lead to a good data
collapse, are consistent with standard percolation exponents.  Notice
that, in contrast to the findings in the $|\phi|^4$ theory, the vortices
appear to proliferate in the confinement phase \textit{after} passing
through the crossover region ($\kappa_\mathrm{per}<\kappa_\times$).
We expect this to be related to the presence of monopoles which can act
as sources for the vortices.  To assess this, we study in the next
section how the relative positions of the percolation threshold and the
crossover region vary with changes in the parameters $\lambda$ and
$\beta$.

\section{Fine Tuning}
\label{sec:threshold}
Since the vertex percolation threshold $\kappa_\mathrm{per}$ and the
phase boundary $\kappa_\mathrm{c}$ coincide in the first-order
transition region, it is expected that as we approach this region, the
discrepancy found in the previous section at $\beta=1.1$ and
$\lambda=0.2$ becomes smaller.  To verify this, we repeat our analysis
at $\beta=1.1$ and $\lambda=0.075$, which is closer to the first-order
transition region.  We find that the difference between the locations of
the percolation threshold and the crossover determined using the overall
vortex line susceptibility $\chi(v)$ indeed becomes smaller, changing
from about $0.018$ at $\lambda=0.2$ to $0.007$ here.  Moving in the
opposite direction of increasing $\lambda$, we observe the discrepancy
to increase, becoming as large as about $0.1$ in the London limit
$\lambda \to \infty$.  The percolation threshold still appears in the
confinement phase after passing through the crossover region.  That is,
changes in the Higgs self-coupling $\lambda$ seem to leave the relative
positions of the crossover and the apparent vortex proliferation
threshold unchanged.

We next vary $\beta$ and set $\beta=2.0$ and $\lambda=0.2$.  By
increasing $\beta$, one suppresses the monopoles.  They completely
disappear in the limit $\beta \to \infty$, where the theory looses its
compactness. Unfortunately, simulations at $\beta=2.0, \lambda=0.2$ are
computationally much more challenging than at $\beta=1.1, \lambda=0.2$
as autocorrelation times are much longer.  We therefore restrict
ourselves to lattice sizes up to $L=36$. Moreover, observables other
than percolation observables become less useful as can be seen from our
example in Fig.~\ref{fig:compare}.  Whereas the susceptibility of the
percolation probability has a pronounced peak and a clear maximum, the
maximum in the coslink susceptibility $\chi(C)$ would be difficult to
identify without reweighting.  The large error bars obtained for the
peak location of $\chi(C)$ reflect the absence of a pronounced peak.

\begin{figure}[t]
  \centering \includegraphics[width=0.8\textwidth]{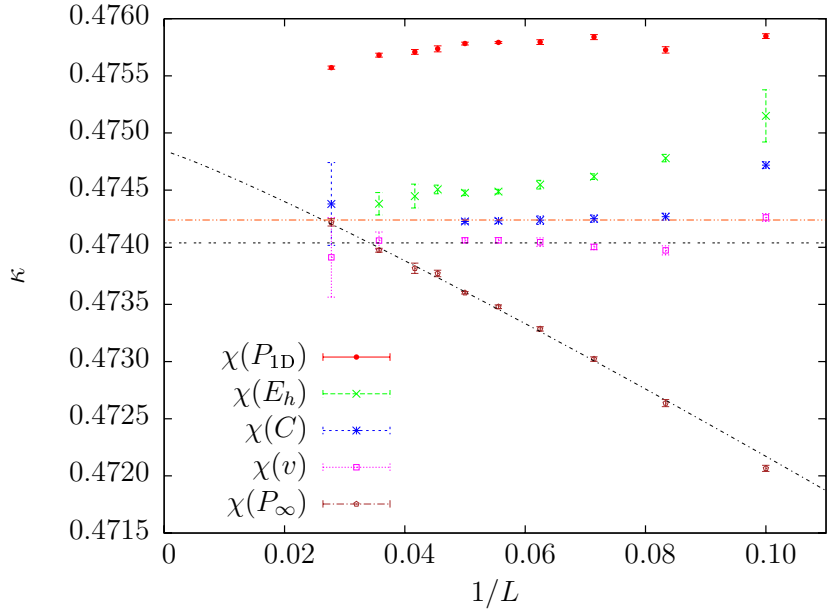}
  \caption{\label{fig:scalingbeta2.0} Locations of the susceptibility
    maxima of various observables as a function of $1/L$ for
    $\beta=2.0$.}
\end{figure}

Our main conclusion of the simulations at $\beta=2.0$ and $\lambda=0.2$
is that the locations of the apparent percolation threshold and the
crossover determined using the overall vortex line susceptibility
$\chi(v)$ have changed relative positions.  This conclusion is based on
Fig.~\ref{fig:scalingbeta2.0} in which $\kappa_\mathrm{per}(L)$,
estimated using different percolation observables, is plotted as a
function of $1/L$ to see their tendency for $L \to \infty$.  The first
observation is that the data points are much closer to each other than
was the case at $\beta=1.1$ (see Fig.~\ref{fig:scaling_sus1}).  As
before, $\kappa_\mathrm{per}(L)$ obtained from the susceptibility of the
percolation strength (lowest set of data points) shows the largest
corrections but increases monotonically for $L\to\infty$.  The values
$\kappa_\mathrm{per}(L)$ obtained from the percolation probability
$P_\mathrm{1D}$ show less drastic corrections but are more difficult to
extrapolate to the infinite-volume limit.
The two sets of data points both extrapolate to a value around
$\kappa_\mathrm{per} \approx 0.4748$ which follows from fitting
$\chi(P_\infty)$ to Eq.~(\ref{eqn:peakscaling}).

Figure \ref{fig:scalingbeta2.0} displays in addition to the locations of
the susceptibility maxima of the vortex line density, also those of the
coslink energy and the hopping energy. For $L>16$ the heights and
locations of these susceptibilities remain constant, so that, as far as
these observables are concerned, the infinite-volume limit is reached.
The location of the crossover in the infinite-volume limit determined
using these observables is below all estimates of the percolation
threshold.  In other words, while for $\beta=1.1, \lambda=0.2$,
$\kappa_\mathrm{per} < \kappa_\times$, here the relative positions
have changed, $\kappa_\mathrm{per} > \kappa_\times$, and the
vortices proliferate already in the Higgs phase before entering the
crossover region.  By continuity we then expect at some value of
$\beta$ in the interval $1.1 < \beta <2.0$ the percolation threshold of
the vortices to coincide with the location of the crossover.  Since
varying $\beta$ physically changes the monopole density, it is tempting
to conclude that monopoles impede the formation of percolating vortex
lines, and that by adjusting the monopole density, the location of the
apparent vortex percolation threshold can be fine-tuned to coincide with
that of the crossover.  The use of the word ``apparent'' here is to
underscore that vortex networks identified with our tracing rules are
only an approximation to the true networks.  The element of impediment
introduced by the monopoles possibly plays a similar role as the
stochastic element in the Fortuin-Kasteleyn construction of spin clusters
in the Potts model \cite{FrotuinKasteleyn}.

We finally estimate the critical exponent $\nu_\mathrm{per}$ for
$\beta=2.0$. Using the percolation probability as input observable for
our collapsing routine, we arrive at the value
$\nu_\mathrm{per}=0.88(1)$ and $\kappa_\mathrm{per}=0.4748(1)$.  As a
crosscheck we fit the scaling of $\chi(P_{\infty})$ with the lattice
size $L$ to Eq.~(\ref{eqn:peakscaling}) giving
$\nu_\mathrm{per}=0.88(8)$ and $\kappa_\mathrm{per}=0.47484(15)$ with a
$\chi^2/\mathrm{DOF}=2.1$.  This is again consistent with the standard
percolation exponent. Although the relative positions of the apparent
percolation threshold and the crossover have changed, variations in the
inverse gauge coupling parameter $\beta$ appear not to change the value
of this critical exponent.

\section{Conclusions}
\label{sec:conclusions}
The vortices arising in the compact Abelian Higgs model have been
investigated by means of Monte Carlo simulations on a cubic lattice, and
analyzed with the help of observables known from percolation theory.
Because their behavior is more pronounced, percolation observables are
better suited than other observables to probe the phase boundary, both
in the first-order transition region as well as in the crossover region
of the phase diagram.  In the region where the Higgs and confinement
phases are separated by a first-order transition, the vortices percolate
right at the phase boundary.  Since the rules applied to trace out the
vortices result in networks that are in general only an approximation to
the true ones, it is concluded that the discontinous first-order
transition is forgivable of the resulting inaccuracies.  The vortex
network reflects the first-order nature of the transition in this region
of the phase diagram.  In the crossover region, the vortices still
percolate.  The percolation observables show second-order critical
behavior along the Kert\'esz line that is characterized by the usual
percolation exponents.  The location of the vortex percolation threshold
estimated using percolation observables does not coincide with that of
the crossover estimated using the vortex line density.  Ideally, one
would expect both to coincide.  Also the $|\phi|^4$ theory in 3D, which
undergoes a continuous phase transition, shows a similar behavior
\cite{bittner3}.  Whereas the estimate based on the line density
coincides with the critical temperature, the estimates based on any of
the percolation observables considered do not agree within error
bars. This discrepancy may arise because the vortex networks are not
correctly traced out or because a stochastic or impeding element in the
construction of a network is missing.  The monopoles appear to play such
a role.

\section{Acknowledgements}
We wish to thank Arwed Schiller for useful discussions at an early stage
of this project.  S.W. acknowledges a PhD fellowship from the
Studienstiftung des\break deutschen Volkes.  A.S. is indebted to Professor H.
Kleinert for the kind hospitality at the Freie Universit\"at Berlin.
This work was partially supported by the Deutsche Forschungsgemeinschaft
under grant Nos.  JA483/22-1 and 23-1, a computer time grant on the JUMP
computer of NIC at Forschungszentrum J\"ulich under project number
HLZ12, and the EC Marie Curie Research and Training Network ENRAGE 
``Random Geometry and Random Matrices: From Quantum Gravity to
Econophysics'', under grant No.~MRTN-CT-2004-005616.

\bibliographystyle{prsty} 
\bibliography{literature_1}

\end{document}